\documentclass[12pt]{article}
\usepackage{comment}
\usepackage{enumerate}
\usepackage{mathrsfs}
\usepackage{units}
\usepackage{hyperref,graphicx,amsfonts,amssymb,amsthm,amsmath,psfrag}
\hypersetup{pageanchor=false,citecolor=blue,urlcolor=blue}
\usepackage[toc,page]{appendix}
\usepackage{subfigure}
\usepackage{slashed}
\usepackage{tikz}
\usepackage{booktabs}
\usepackage{siunitx}
\usepackage{cite}
\usepackage{amssymb}
\usepackage{physics}

\usepackage[normalem]{ulem}	
\newcommand\redout{\bgroup\markoverwith
{\textcolor{red}{\rule[.5ex]{8pt}{0.8pt}}}\ULon}
\usepackage{xcolor}
\usepackage{color}
\usepackage{todonotes}

\usetikzlibrary{decorations.markings,snakes}
\tikzset{
  fermionline/.style={line width=1pt,postaction={decorate},
    decoration={markings,
      mark=at position 0.5 with {\draw[-stealth] (0,0)--(2pt,0);}}},
  bosonline/.style={line width=1pt,decorate,
    decoration={snake,amplitude=1,segment length=4}},
  higgsline/.style={line width=1pt,dashed}
}

\usepackage{amsmath,amssymb,exscale}
\usepackage{graphicx}
\usepackage{epsfig}
\usepackage{multicol}
\usepackage{color}
\usepackage{mathrsfs}
\usepackage{blindtext}
 \usepackage{fancyhdr}
\usepackage{hyperref}
\usepackage{cite}
\usepackage{mathtools}
\usepackage{amsmath}
\usepackage{rotating,slashed,amsmath,charter,xcolor,catchfilebetweentags,ifluatex}

\usepackage{graphicx}
\usepackage{sidecap}

\usepackage[latin1]{inputenc} 
\usepackage{multicol}  
 
 \usepackage{titlesec}
 
 \usepackage{rotating,slashed,xcolor,amsfonts,expdlist,charter}

\numberwithin{equation}{section}

\usepackage{xcolor}
\usepackage{sectsty}
\usepackage{mdframed}
\usepackage{titletoc}

\definecolor{secnum}{RGB}{13,151,225}
\definecolor{ptcbackground}{RGB}{212,237,252}
\definecolor{ptctitle}{RGB}{0,177,235}

\titlecontents{lsection}
  [5.8em]{\sffamily}
  {\color{secnum}\contentslabel{2.3em}\normalcolor}{}
  {\titlerule*[1000pc]{.}\contentspage\\\hspace*{-5.8em}\vspace*{5pt}%
    \color{white}\rule{\dimexpr\textwidth-15.5pt\relax}{1pt}}

\newcommand{\be}{\begin{equation} }
\newcommand{\ee}{\end{equation}}
\newcommand{\cL}{\mathcal{L}}

\newcommand{\MSb}{$\overline{\text{MS}}$}

\DeclareMathOperator{\arcsinh}{arcsinh}

\RequirePackage{etex}

\usepackage{hyperref}
\hypersetup{colorlinks,bookmarksopen,bookmarksnumbered,citecolor=blue,
linkcolor=blue,pdfstartview=FitH,urlcolor=blue}
\usepackage{slashed}

\definecolor{blus}{cmyk}{1,0.9,0,0.1}
\definecolor{verdes}{cmyk}{0.99,0,0.59,0.65}
\definecolor{rossos}{cmyk}{0,1,1,0.55}
\definecolor{redy}{cmyk}{0,1,1,0.7}
\definecolor{greeny}{cmyk}{0.99,0,0.59,0.98}
\definecolor{green-go}{cmyk}{0.79,0,0.59,0.5}

\usepackage{titlesec}

\newcommand{\beq}{\begin{equation}}
\newcommand{\eeq}{\end{equation}}

\newcommand{\tmtextbf}[1]{{\bfseries{#1}}}
\newcommand{\tmtextrm}[1]{{\rmfamily{#1}}}
\def\be{\begin{equation}}
\def\ee{\end{equation}}
\def\ba{\begin{array} }
\def\bac{\begin{array} {c}}
\def\bacc{\begin{array} {cc}}
\def\baccc{\begin{array} {ccc}}
\def\bacccc{\begin{array} {cccc}}
\def\ea{\end{array}}
\def\bea{\begin{eqnarray}}
\def\eea{\end{eqnarray}}

\definecolor{red}{rgb}{1,0,0}

\def\psl{\hbox{\hbox{${p}$}}\kern-1.9mm{\hbox{${/}$}}}
\def\dsl{\hbox{\hbox{${\partial}$}}\kern-2.2mm{\hbox{${/}$}}}
\def\Dsl{\hbox{\hbox{${D}$}}\kern-2.6mm{\hbox{${/}$}}}

\newcommand{\gappeq}{{\rlap{{\raise}.5ex\text{\ensuremath{>}}}{{\lower}.5ex\text{\ensuremath{\sim}}}}}
\newcommand{\lappeq}{{\rlap{{\raise}.5ex\text{\ensuremath{<}}}{{\lower}.5ex\text{\ensuremath{\sim}}}}}
\newcommand{\I}{\tmtextrm{1{\kern}-.24em l}}

\begin{document}
\topmargin -1.0cm
\oddsidemargin 0.9cm
\evensidemargin -0.5cm

{\vspace{-1cm}}
\begin{center}

\vspace{-1cm}

\textcolor{blue}{{\Huge \tmtextbf{ Higgs inflation}} {\vspace{.5cm}}}\\
 
\vspace{1.9cm}

{\large  {\bf Javier Rubio }
{\em  

\vspace{.4cm}
Institut f\"ur Theoretische Physik, Ruprecht-Karls-Universit\"at Heidelberg, \\
Philosophenweg 16, 69120 Heidelberg, Germany \\ 

\vspace{0.4cm}

\vspace{0.2cm}

 \vspace{0.5cm}
}

\vspace{.3cm}

\vspace{0.5cm}
}
\vspace{0.cm}
\end{center}

\noindent ------------------------------------------------------------------------------------------------------------------------------
%------------

\begin{center}
{\bf \large Abstract}  
\end{center}
The properties of the recently discovered Higgs boson together with the absence of new physics at collider experiments allows us to speculate about consistently extending the Standard Model of particle physics all the way up to the Planck scale. In this context, the Standard Model Higgs non-minimally coupled to gravity could be responsible for the symmetry properties of the Universe at large scales and for the generation of the primordial spectrum of curvature perturbations seeding structure formation. We overview the minimalistic Higgs inflation scenario, its predictions, open issues and extensions and discuss its interplay with the possible metastability of the Standard Model vacuum.

  \vspace{0.9cm}  
\noindent ------------------------------------------------------------------------------------------------------------------------------
%------------

\vspace{6cm}

\noindent Email: j.rubio@thphys.uni-heidelberg.de

\newpage
\tableofcontents
\newpage

\section{Introduction and summary} 

Inflation is nowadays a well-established paradigm \cite{Starobinsky:1980te, Guth:1980zm, Mukhanov:1981xt, Linde:1981mu, Albrecht:1982wi,Linde:1983gd} able to explain the flatness, homogeneity and isotropy of the Universe and the generation of the primordial density fluctuations seeding structure formation \cite{Hawking:1982cz,Starobinsky:1982ee, Sasaki:1986hm, Mukhanov:1988jd}. In spite of the phenomenological success, the inflaton's nature  remains unknown and its role could be played by any particle physics candidate able to imitate a slowly-moving scalar field in the very early Universe. 

\begin{figure}
\centering
\includegraphics[width=0.6\textwidth]{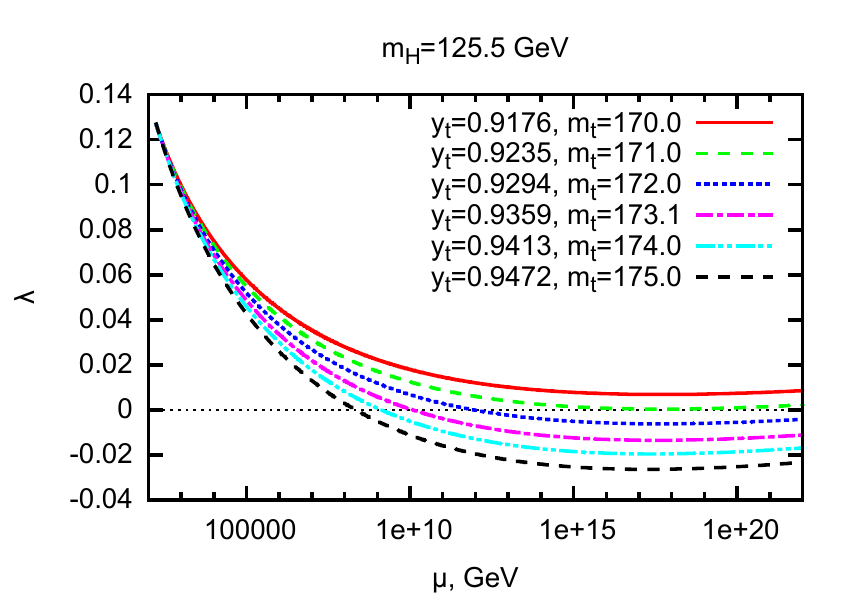}
\caption{The running of the Higgs self-coupling following from the Standard Model renormalization group equations  for several values of the top quark Yukawa coupling at the electroweak scale and a fixed Higgs boson mass $m_H= 125.5$\, GeV \cite{Bezrukov:2014ipa}.}\label{fig:running}
\end{figure}

\begin{figure}
\centering
\includegraphics[width=0.65\textwidth]{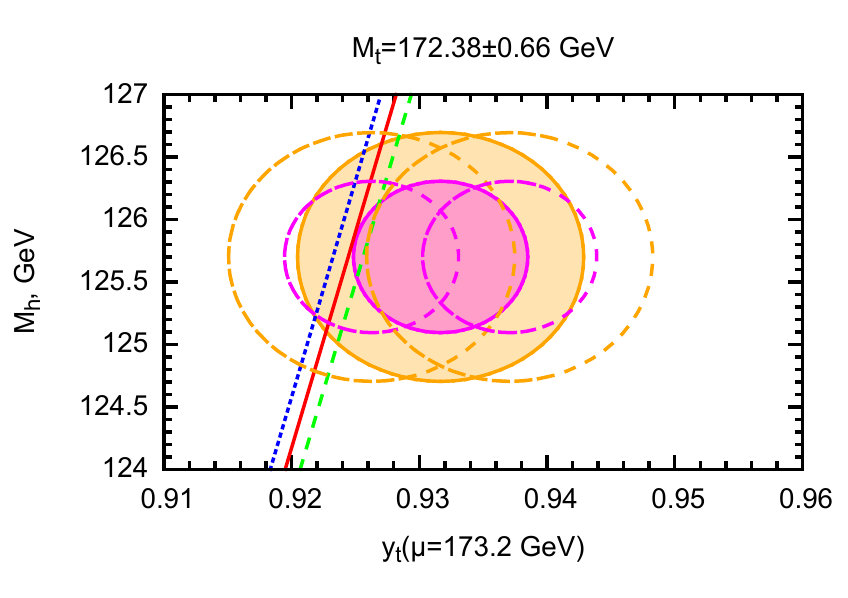}
\caption{The SM stability and metastability regions for a renormalization point $\mu=173.2$ GeV  in the \MSb \,\,scheme  \cite{Bezrukov:2014ina}. The solid red line corresponds to the critical top quark Yukawa coupling \eqref{eq:ycrit} leading to vacuum instability at a sub-Planckian energy scale $\mu_0$, with the dashed lines accounting for uncertainties associated with the strong coupling constant. To the left (right) of these diagonal lines, the SM vacuum is unstable (metastable). The filled elliptical contours account for the $1\sigma$ and $2\sigma$ experimental errors on the Higgs mass in Ref.~\cite{Agashe:2014kda} and the CMS (Monte-Carlo) top quark mass in Ref.~\cite{CMS:2014hta}, namely $m_H=125.7 \pm 0.4$ GeV and  $m_t = 172.38 \pm 0.10\, {\rm (stat)} \pm 0.66\, {\rm(syst)}\, {\rm GeV}$ (note that the \textit{current} value of the Higgs mass is slightly lower \cite{Tanabashi:2018oca}). The additional empty contours illustrate the shifts associated with the theoretically ambiguous relation between the top quark Yukawa coupling and the (Monte-Carlo) top quark mass (cf.~ Ref.~ \cite{Bezrukov:2014ina} for details).}\label{fig:CMS}
\end{figure}

In spite of dedicated searches, the only outcome of the Large Hadron Collider (LHC) experiments till date is a scalar particle with mass \cite{Aad:2012tfa,Chatrchyan:2012xdj,Tanabashi:2018oca}
\begin{equation} \label{eq:Hmass}
m_{H}=125.18 \pm  0.16 \hspace{2mm} {\rm GeV}
\end{equation}
 and properties similar to those of the Standard Model (SM) Higgs.  The mass value \eqref{eq:Hmass} is certainly particular since it allows to extend the SM up to the Planck scale without leaving the perturbative regime \cite{Shaposhnikov:2009pv}. The main limitation to this appealing scenario is the potential  instability of the electroweak vacuum at high energies. Roughly speaking, the value of the Higgs self-coupling following from the SM renormalization group equations decreases with energy till a given scale and starts increasing thereafter. Whether it stays positive all the way up to the Planck scale, or turns negative at some intermediate scale $\mu_0$ depends, mainly, on the interplay between the Higgs mass $m_H$ and the top quark Yukawa coupling $y_t$ extracted from the reconstructed Monte-Carlo top mass in collider experiments \cite{Butenschoen:2016lpz}, cf.~Fig.~\ref{fig:running}. Neglecting the effect of gravitational corrections, the critical value $y_t^{\rm crit}$ separating the region of absolute stability from the metastability/instability\footnote{The metastability region is defined as the parameter space leading to vacuum instability at energies below the Planck scale but with an electroweak vacuum lifetime longer than the age of the Universe.} regions is given by \cite{Bezrukov:2014ina}
\begin{equation}\label{eq:ycrit}
y_t^{\rm crit}=0.9244\pm 0.0012 \,\frac{m_H/{\rm GeV}-125.7}{0.4}+0.0012\,\frac{\alpha_s(m_Z)-0.01184}{0.0007}\,,
\end{equation}
 with $\alpha_s(m_Z)$ the strong coupling constant at the $Z$ boson mass. Within the present experimental and theoretical uncertainties \cite{Bezrukov:2014ina,Butenschoen:2016lpz,Espinosa:2016nld}, the SM is compatible with vacuum instability, metastability and absolute stability \cite{Bezrukov:2012sa,Degrassi:2012ry,Buttazzo:2013uya,Espinosa:2015qea,Espinosa:2015kwx}, with the separation among the different cases strongly depending on the ultraviolet completion of gravity ~\cite{Branchina:2013jra,Branchina:2014usa,Branchina:2014rva}, cf.~Fig.~\ref{fig:CMS}. 

 In the absence of physics beyond the SM, it is certainly tempting to identify the recently discovered Higgs boson with the inflaton condensate. Unfortunately, the Higgs self-interaction significantly  exceeds the value $\lambda \sim 10^{-13}$ leading to a sufficiently long inflationary period without generating an excessively large spectrum of primordial density perturbations \cite{Linde:1983gd}. The 
situation is unchanged if one considers the renormalization group enhanced potential following from the extrapolation of the SM renormalization group equations to the inflationary scale  \cite{Isidori:2007vm, Hamada:2013mya, Fairbairn:2014nxa}. The simplest way out is to modify the Higgs field kinetic term in the large-field regime. In Higgs inflation\footnote{An alternative possibility involving a derivative coupling among the Einstein tensor and the Higgs kinetic term was considered in Refs.~\cite{Germani:2010gm,Germani:2010ux,Fumagalli:2017cdo}.} this is done by including a direct coupling to the Ricci scalar $R$ \cite{Bezrukov:2007ep}, namely\footnote{ Prior to Ref.~\cite{Bezrukov:2007ep}, the effect of non-minimal couplings had been extensively studied in the literature (see for instance Refs.~\cite{Minkowski:1977aj,Zee:1978wi,Smolin:1979uz,Spokoiny:1984bd,Futamase:1987ua,Salopek:1988qh,Fakir:1990iu,Fakir:1990eg,Makino:1991sg,Fakir:1992cg,Kaiser:1994vs,CervantesCota:1994zf,CervantesCota:1995tz,Komatsu:1997hv}), but never with the SM Higgs as the inflaton.}
\begin{equation}\label{deltaS}
\delta S=\int d^4 x\sqrt{-g} \left[ \xi H^\dagger H R\right] \,,
\end{equation}
with $H$ the Higgs doublet and $\xi$ a dimensionless constant to be fixed by observations. The inclusion of the non-minimal coupling \eqref{deltaS} can be understood as an inevitable consequence of the quantization of the SM in a gravitational background, where this term is indeed required for the renormalization of the energy-momentum tensor \cite{Callan:1970ze,Birrell:1982ix}. 
 %%%%%%%%%%%%%%%%%%%%%%%%%%%%%%%  
%%%%%%%%%%%%%%%%%%%%%%%%%%%%%%%  
%%%%%%%%%%%%%%%%%%%%%%%%%%%%%%%  

When written in the Einstein frame, the Higgs inflation scenario displays two distinct regimes. At low energies, it  approximately coincides  with the SM minimally coupled to gravity. At high energies, it becomes a chiral SM with no radial Higgs component \cite{Bezrukov:2009db,Dutta:2007st}. In this latter regime, the effective Higgs potential becomes exponentially flat, allowing for inflation with the usual chaotic initial conditions. The associated predictions depend only on the number of e-folds of inflation, which is itself related to the duration of the heating stage. As the type and strength of the interactions among the Higgs field and other SM particles is experimentally known, the duration of this entropy production era can be computed in detail ~\cite{GarciaBellido:2008ab,Bezrukov:2008ut,Bezrukov:2011sz,Repond:2016sol,Ema:2016dny}, leading to precise inflationary predictions in perfect agreement with observations \cite{Akrami:2018odb}. 
 
The situation becomes more complicated when quantum corrections are included. The shape of the inflationary potential depends then on the values of the Higgs mass and top Yukawa coupling \textit{at the inflationary scale}. In addition to the plateau already existing at tree-level \cite{Bezrukov:2014bra,Fumagalli:2016lls,Fumagalli:2016sof,Enckell:2016xse}, the renormalization group enhanced potential can develop a quasi-inflection point along the inflationary trajectory \cite{Allison:2013uaa, Bezrukov:2014bra, Hamada:2014iga, Bezrukov:2014ipa, Rubio:2015zia, Fumagalli:2016lls, Enckell:2016xse, Bezrukov:2017dyv, Rasanen:2017ivk} or a hilltop \cite{Fumagalli:2016lls, Rasanen:2017ivk,Enckell:2018kkc}, with different cases giving rise to different predictions. 

 Although a precise measurement of the inflationary observables could be understood as an interesting consistency check between cosmological observations and  particle physics experiments  \cite{Espinosa:2007qp,Barvinsky:2008ia,DeSimone:2008ei,Bezrukov:2008ej,Bezrukov:2009db,Barvinsky:2009ii,Popa:2010xc,Bezrukov:2012sa,Salvio:2013rja}, the low- to high-energy connection is subject to unavoidable ambiguities related to the non-renormalizability of the model \cite{Barbon:2009ya,Burgess:2009ea,Burgess:2010zq,Bezrukov:2010jz,George:2013iia,Bezrukov:2014ipa, Rubio:2015zia, George:2015nza, Fumagalli:2016lls, Enckell:2016xse, Bezrukov:2017dyv}. In particular, the finite parts of the counterterms needed to renormalize the tree-level action lead to localized jumps in the SM renormalization group equations when connected to the chiral phase of Higgs inflation. The strength of these jumps encodes the remnants of the ultraviolet completion and cannot be determined within effective field theory approach \cite{Bezrukov:2010jz,Bezrukov:2014ipa,Burgess:2014lza}.
If the finite parts are significantly smaller than the associated coupling constants, Higgs inflation leads to a direct connection among the SM parameters measured at collider experiments and the large scale properties of the Universe, provided that the former do not give rise to vacuum instability. On the contrary, if the jumps in the coupling constants are large, the relation between high- and low-energy physics is lost, but Higgs inflation can surprisingly occur even if the electroweak vacuum is not completely stable \cite{Bezrukov:2014ipa}. 

In this article we review the minimalistic Higgs inflation scenario, its predictions, open issues and extensions, and discuss its interplay with the potential metastability of the SM vacuum. The paper is organized as follows: 

\begin{itemize}
\item The general framework is introduced in Section \ref{sec:HItree}. To illustrate the effect of  non-minimal couplings, we consider an induced gravity scenario in which the effective Newton constant is completely determined by the Higgs vacuum expectation value. Having this toy model in mind, we reformulate Higgs inflation in the so-called Einstein frame in which the coupling to gravity is minimal and all non-linearities appear in the scalar sector of the theory. After emphasizing the pole structure of the Einstein-frame kinetic term and its role in the asymptotic flatness of the Higgs inflation potential, we compute the tree-level inflationary observables and discuss the decoupling properties of the SM degrees of freedom.   

\item The limitations of Higgs inflation as a fundamental theory are reviewed in 
Section \ref{sec:EFT}. In particular, we present a detailed derivation of the cutoff scales signaling the violation of perturbative unitarity in different scattering processes and advocate the interpretation of Higgs inflation as an effective field theory to be supplemented by an infinite set of higher dimensional operators. Afterwards, we adopt a self-consistent approach to Higgs inflation and formulate the set of assumptions leading to a controllable relation between low- and high-energy observables. Based on the resulting framework, we analyze the contribution of quantum corrections to the renormalization group enhanced potential and their impact on the inflationary observables. We finish this section by discussing the potential issues of Higgs inflation with the metastability of the SM vacuum.

\item Some extensions and alternatives to the simplest Higgs inflation scenario are considered in Section \ref{sec:variations}. In particular, we address the difference between the metric and Palatini formulations of the theory and its extension to a fully scale invariant framework ~\cite{Shaposhnikov:2008xb,GarciaBellido:2011de,Blas:2011ac,Bezrukov:2012hx,GarciaBellido:2012zu,Rubio:2014wta,Trashorras:2016azl,Karananas:2016kyt,Ferreira:2016vsc,Ferreira:2016wem,Ferreira:2016kxi,Casas:2017wjh,Ferreira:2018qss,Ferreira:2018itt,Casas:2018fum}. The inflationary predictions in these models are put in one to one correspondence with the pole structure of the Einstein-frame kinetic term, allowing for an easy comparison with the results of the standard Higgs inflation scenario.  
\end{itemize}

Overall, we intend to complement the existing monographs in the literature \cite{Bezrukov:2013fka,Bezrukov:2015dty,Moss:2015fma} by i) providing a further insight on the classical formulation of Higgs inflation and by ii) focusing on the uncertainties associated with the non-renormalizability of the theory and their impact on model building.
 
\section{General framework}\label{sec:HItree}  

The inflationary paradigm is usually  formulated in terms of conditions on the local flatness on an arbitrary potential, which can in principle contain a large number of extrema and slopes \cite{Artymowski:2016pjz}. This flatness is usually related to the existence of some approximate shift-symmetry, which, for the purposes of Higgs inflation, is convenient to reformulate as a non-linear realization of approximate scale-invariance. 
\subsection{Induced gravity}\label{sec:IG}

Let us start by considering an \textit{induced gravity} scenario
  \begin{equation}
\label{lagrIG}
S = \int d^4x \sqrt{-g} \left[ \frac{\xi h^2}{2}   R -\frac{1}{2}\left(\partial h\right)^2- \frac{\lambda}{4}h^4-\frac{1}{4}{\cal F}_{\mu\nu}{\cal F}^{\mu\nu}-\frac{g^2}{4} \, h^2 B_\mu B^\mu - i\bar\psi \slashed\partial\psi - \frac{y}{\sqrt{2}} h \bar\psi \psi \right]\,,
\end{equation} 
involving a scalar field $h$, a vector field $B_\mu$ and a fermion field $\psi$, with interactions similar to those appearing in the SM of particle physics when  written in the unitary gauge $H=(0, h/\sqrt{2})^T$. The quantity ${\cal F}_{\mu\nu}{\cal F}^{\mu\nu}$ stands for the standard $B_\mu$ kinetic term, which for simplicity we take to be Abelian. In this toy model, the effective Newton constant is induced by the scalar field expectation value,
\be 
G_{N,\rm eff}\equiv \frac{1}{8\pi\xi h^2}\,. 
\ee 
In order for $G_{N,\rm eff}$ to be well-behaved, the non-minimal coupling $\xi$ is restricted to take positive values. This condition is equivalent to require the semi-positive definiteness of the scalar field kinetic term, as can be easily seen by performing a field redefinition $h^2\to h^2/\xi$. 

An important property of the \textit{induced gravity} action \eqref{lagrIG} is its invariance under scale transformations
\be \label{eq:dilat}
x^\mu\to \bar x^{\mu}=\alpha \,x^\mu\,,  \hspace{20mm} \varphi(x)\to
\bar \varphi(\bar x)=\alpha^{\Delta_\varphi} \varphi(\alpha\, x) \,.
\ee
Here $\alpha$ is an arbitrary constant, $\varphi(x)$ compactly denotes the various fields in the model and $\Delta_\varphi$'s are their corresponding scaling dimensions.   
The consequences of this dilatation symmetry are more easily understood in a minimally-coupled frame displaying the standard Einstein-Hilbert term. This \textit{Einstein frame} is achieved by a Weyl redefinition of the metric\footnote{In spite of its extensive use in the literature, we avoid referring to point-wise rescalings of the metric  as ``conformal transformations.'' For a comprehensive discussion on the differences between Weyl and conformal symmetries, see for instance Refs.~\cite{Karananas:2015ioa,Karananas:2015eha}.}
\be \label{eq:weyltransformation}
g_{\mu\nu}\rightarrow \Theta \, g_{\mu\nu}\,, \hspace{30mm}\Theta \equiv \frac{F^2_\infty}{h^2} \,, \hspace{30mm} F_\infty\equiv \frac{M_P}{\sqrt{\xi}}\,,
\ee 
together with a Weyl rescaling of the vector and fermion fields,
\begin{equation}
A_\mu \to \Theta^{-1/2} \, A_\mu \,,  \hspace{20mm} \psi\to \Theta^{-3/4}\, \psi \,. 
\end{equation}
After some trivial algebra we obtain an Einstein-frame action\footnote{In order to keep the notation as simple as possible, we will not introduce different notations for the quantities defined in different Weyl-related frames. In particular, the implicit Lorentz contractions in this article should be understood to be performed  with the metric of the frame under consideration.}
\be\label{EframeIG0}
S = \int d^4x \sqrt{-g} \left[\frac{M_P^2}{2}R 
-\frac{1}{2}M_P^2K(\Theta)(\partial \Theta)^2-\frac{\lambda}{4}F_\infty^4-\frac{1}{4}{\cal F}_{\mu\nu}{\cal F}^{\mu\nu}-\frac{g^2}{4} \, F_\infty^2 B_\mu B^\mu - i\bar\psi \slashed\partial\psi - \frac{y}{\sqrt{2}}F_\infty \bar\psi \psi\right] \,,
\ee
containing a non-canonical term for the $\Theta$ field. The coefficient of this kinetic term,
\be \label{eq:KIG}
K(\Theta)\equiv\frac{1}{4\vert a \vert\, \Theta^2}\,,
\ee
involves a quadratic pole at $\Theta=0$ and a constant
\be\label{ainf}
a  \equiv -\frac{\xi}{1+6\xi}<0\,,
\ee
varying between zero at $\xi=0$ and $-1/6$ when $\xi\to\infty$. The $\Theta$-field kinetic term can be made canonical by performing an additional field redefinition,
\begin{equation}\label{eq:expmap}
\Theta^{-1}=\exp\left({\frac{2\sqrt{\vert a\vert}\phi}{M_P}} \right)\,,
%\phi \equiv -\frac{M_P}{2\sqrt{\vert a_c\vert}}\log \left(\frac{M_P^2}{\xi h^2} \right) 
\end{equation} 
mapping the vicinity of the pole at $\Theta=0$ to $\phi\to\infty$. The resulting action 
\be\label{EframeIG}
S = \int d^4x \sqrt{-g} \left[\frac{M_P^2}{2}R 
-\frac{1}{2}(\partial \phi)^2-\frac{\lambda}{4}F_\infty^4-\frac{1}{4}{\cal F}_{\mu\nu}{\cal F}^{\mu\nu}-\frac{g^2}{4} \, F_\infty^2 B_\mu B^\mu - i\bar\psi \slashed\partial\psi + \frac{y}{\sqrt{2}}F_\infty \bar\psi \psi\right] \,,
\ee
is invariant under shift transformations $\phi \to \phi+C$, with $C$ a constant. The exponential mapping in Eq.~\eqref{eq:expmap} indicates that such translational symmetry is nothing else than the non-linear realization of the original scale invariance we started with in Eq.~\eqref{lagrIG}  \cite{Csaki:2014bua}. 
The Einstein-frame transition in Eq.~\eqref{eq:weyltransformation} is indeed equivalent to the spontaneous breaking of dilatations, since we implicitly required the field $h$ to acquire a non-zero expectation value. The canonically normalized scalar field $\phi$ is the associated Goldstone boson and as such it is completely decoupled from the matter fields $B_\mu$ and $\psi$. The non-minimal coupling to gravity  effectively replaces $h$ by $F_\infty$ in all dimension-4 interactions involving conformal degrees of freedom. Note, however, that this decoupling statement does not apply to scale-invariant extensions including additional scalar fields \cite{Kaiser:2010ps,Kaiser:2012ak,Kaiser:2013sna,GarciaBellido:2011de,Bezrukov:2012hx,Karananas:2016kyt} or other non-conformal interactions such as $R^2$ terms \cite{Starobinsky:1980te,Gorbunov:2010bn,Gorbunov:2012ij,Gorbunov:2012ns}. 

\subsection{Higgs inflation from approximate scale invariance} \label{HI_Jordan}

Although the toy model presented above contains many of the key ingredients of Higgs inflation, it is not phenomenologically viable. In particular, the Einstein-frame potential is completely shift-symmetric and does not allow for inflation to end. On top of this limitation, the scalar field $\phi$ is completely decoupled from all conformal fields, excluding the possibility of entropy production and the eventual onset of a radiation-dominated era. All these phenomenological limitations are intrinsically related to the \textit{exact} realization of scale invariance and as such they should be expected to disappear once a (sizable) dimensionfull parameter is included into the action. This is precisely what happens in Higgs inflation. The total \textit{Higgs inflation} action  \cite{Bezrukov:2007ep}
\begin{equation}\label{SHG}
S= \int d^4x \sqrt{-g} \left[\frac{M^2_P}{2}R+\xi H^\dagger H R+{\cal L}_{\rm SM} \right]\;\,,
\end{equation} 
contains two dimensionfull parameters: the reduced Planck $M_P\equiv 1/\sqrt{8\pi G_N}=2.435 \times 10^{18}$~GeV and the Higgs vacuum expectation value $v_{\rm EW}\simeq 250$ GeV responsible for the masses within the SM Lagrangian density ${\cal L}_{\rm SM}$. Among these two scales, the Planck mass is the most important one at the large field values relevant for inflation. To illustrate how the inclusion of $M_P$ modifies the results of the previous section, let us consider the graviscalar part of Eq.~\eqref{SHG} in the unitary gauge $H=(0, h/\sqrt{2})^T$, namely
  \begin{equation}
\label{lagr}
S = \int d^4x \sqrt{-g} \left[ \frac{M_P^2 + \xi h^2}{2}   R -\frac{1}{2}\left(\partial h\right)^2- U(h) \right]\,,
\end{equation} 
with 
\begin{equation}\label{SBpot}
U(h)=\frac{\lambda}{4}(h^2-v_{\rm EW}^2)^2\,,
\end{equation}
the usual SM symmetry-breaking potential. As in the induced gravity scenario, the inclusion of the non-minimal coupling to gravity changes the strength of the gravitational interaction and makes it dependent on the Higgs field, 
\be \label{GNeff}
G_{N,{\rm eff}}=\frac{G_N}{1+8\pi \xi G_N h^2}\,.
\ee 
In order for the graviton propagator to be well-defined at all $h$ values, the non-minimal coupling $\xi$ must be positive.\footnote{Models with negative $\xi$ have been considered in the literature \cite{Herranen:2014cua, Kamada:2014ufa,Figueroa:2017slm}. In this type of scenarios the gravitational instability at large field values can be avoided by replacing the quadratic coupling $\xi h^2$ by a designed function $\xi f(h)$ remaining  smaller than $M_P^2$ during the whole field regime.} If $\xi\neq 0 $, this requirement translates into a weakening of the effective Newton constant at increasing Higgs values. For non-minimal couplings in the range $
1\ll \xi\ll M_P^2/v^2_{\rm EW}$, this effect is important in the large-field regime $h\gg M_P/\sqrt{\xi}$, but completely negligible otherwise. 

As we did in Section \ref{sec:IG}, it is convenient to reformulate Eq.~\eqref{lagr} in the Einstein frame by performing a Weyl transformation $g_{\mu\nu} \to \Theta \,g_{\mu\nu}$ with 
\begin{equation}\label{Thetadef}
  \Theta^{-1} = 1+\frac{h^2}{F_\infty^2} \,,  \hspace{25mm} F_\infty\equiv \frac{M_P}{\sqrt{\xi}}\,. 
  \end{equation}
In the new frame, all the non-linearities associated with the non-minimal Higgs-gravity interaction are moved to the scalar sector of the theory,
\begin{equation}\label{lagr2}
S= \int d^4x \sqrt{-g}\left[ \frac{M_P^2}{2}   R -\frac{1}{2}M_P^2 K(\Theta)\left(\partial \Theta\right)^2-V(\Theta)\right]\,,
\end{equation}
which contains now a \textit{non-exactly flat} potential
\be \label{potE}
V(\Theta)\equiv U(\Theta)\,\Theta^2=\frac{\lambda F_\infty^4}{4}\left[1-\left(1+\frac{v^2_{\rm EW}}{F^2_\infty}\right)\Theta\right]^2\,,
\ee  
and a non-canonical kinetic sector resulting from the rescaling of the metric determinant and the non-homogeneous part of the Ricci scalar transformation. 
The kinetic function
\begin{equation}\label{K_HI}
  K(\Theta) 
  \equiv \frac{1}{4\vert a\vert \Theta^2}\left(\frac{1-6 \vert a \vert \Theta}{1-\Theta}\right) \,,
%\equiv \frac{1+(\xi+6\xi^2)h^2/M_P^2}{(1+\xi h^2/M_P^2)^2} 
\end{equation}
shares some similarities with that in Eq.~\eqref{eq:KIG}. In particular, it contains two poles located respectively at $\Theta=0$ and $\Theta=1$. The first pole is an \textit{inflationary pole}, like the one appearing in the induced gravity scenario. This pole leads to an enhanced friction for the $\Theta$ field around $\Theta=0$ and allows for inflation to happen even if the potential $V(\Theta)$ is not sufficiently flat.  The second pole is a \textit{Minkowski pole} around which the Weyl transformation equals one and the usual SM action is approximately recovered. To see this explicitly, we carry out an additional field redefinition\,,\footnote{Note that  all equations till this point hold even if the non-minimal coupling $\xi$ is field-dependent \cite{Ezquiaga:2017fvi,Masina:2018ejw}.}
 \begin{equation}
\frac{1}{M^2_P} \left(\frac{d\phi}{d\Theta}\right)^2=K(\Theta)\,,
 \end{equation}
 to recast  Eq.~\eqref{lagr2} in terms of a canonically normalized scalar field $\phi$. This differential equation admits an exact solution \cite{GarciaBellido:2008ab}
 \be \label{phihexact}
\frac{\sqrt{\vert a\vert }\phi}{M_P}=\arcsinh\sqrt{\frac{1-\Theta}{(1-6\vert a\vert )\Theta}}-\sqrt{6\vert a\vert} \arcsinh \sqrt{ \frac{6\vert a\vert (1-\Theta)}{1-6\vert a\vert}      } \,.
 \ee
%  \be 
% \phi = F_\infty \sqrt{1-\Theta^2} F_1\left[\frac{1}{2}; -\frac{1}{2},1; \frac{3}{2}; -6 \xi  \left(1-\Theta^2\right),1-\Theta^2\right]\,,
% \ee
%with $F_1$ the Appell hypergeometric function of two variables.  
%\begin{equation}\label{thetah}
%\Theta^{-1}\simeq \begin{cases} 
%\sech \left(\frac{\sqrt{\vert a\vert}\phi}{M_P}\right) &  \hspace{2mm} \textrm{for} \hspace{2mm} \phi<\phi_{\rm C}\,,  \\
%\exp\left({\frac{\sqrt{\vert a\vert}\phi}{M_P}}\right)& \hspace{2mm} \textrm{for} \hspace{2mm} \phi>\phi_{\rm C}\,,
%\end{cases}
%\end{equation} 
%%%%%%%%%%%%%%%%%%%%%%%%%%%%%%%%%%%%%%%%%%%%%%55
\begin{figure}
\begin{center}
\includegraphics[width=0.5\textwidth]{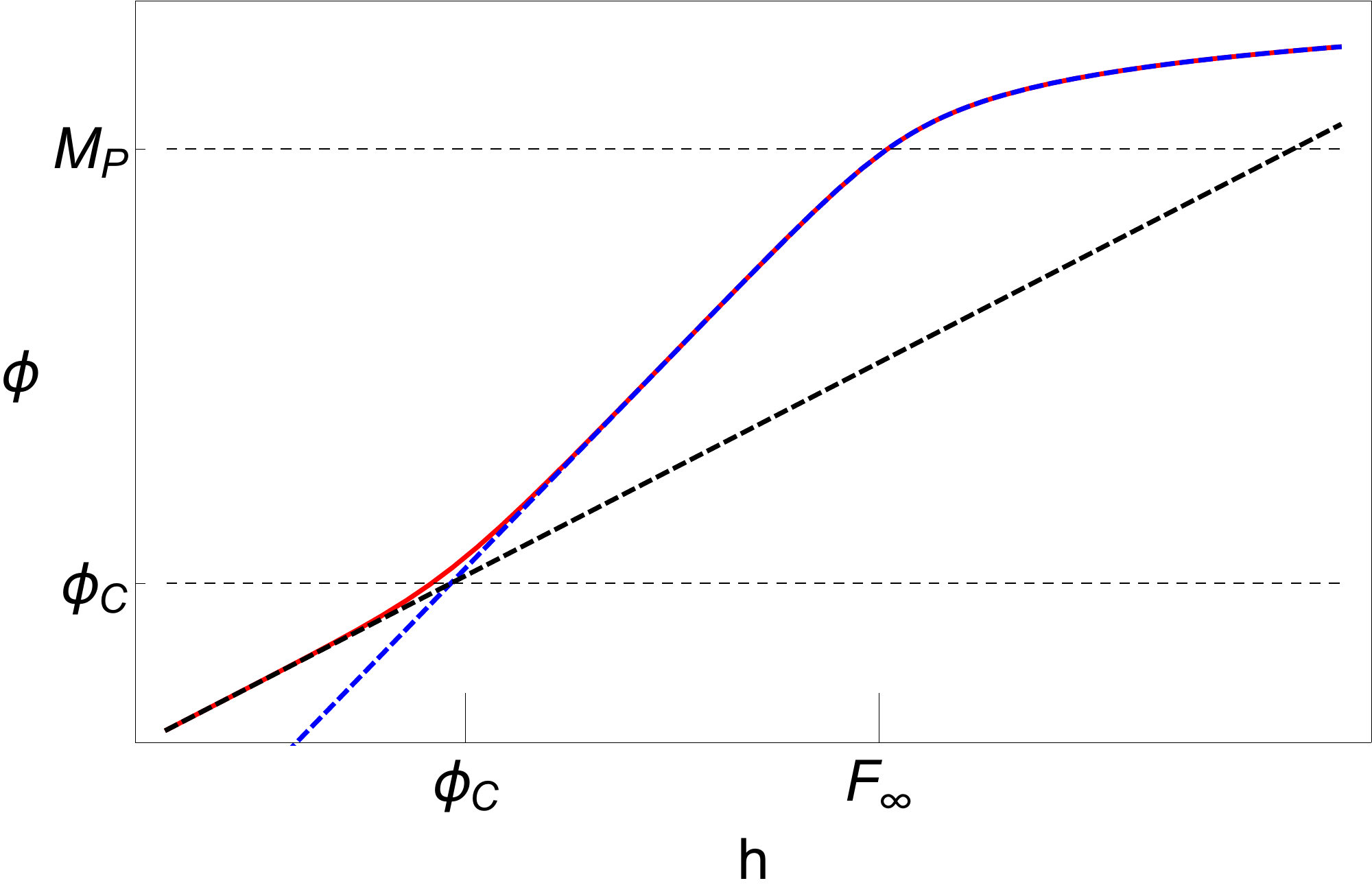}
\caption{Comparison between the approximate expressions in Eq.~\eqref{phih} (dashed black and blue lines) and the exact solution \eqref{phihexact} (solid red). Below the critical scale $\phi_C$, Higgs inflation coincides, up to highly suppressed corrections, with the SM minimally coupled to gravity. Above that scale, the Higgs field starts to decouple from the SM particles. The decoupling becomes efficient at a scale $F_\infty$, beyond which the model can be well approximated by a chiral SM with no radial Higgs component.} \label{fig:comparison}
\end{center}
\end{figure}
%%%%%%%%%%%%%%%%%%%%%%%%%%%%%%%%%%%%%%%%%%%%%%
In terms of the original field $h$, we can distinguish two asymptotic regimes 
\begin{equation}\label{phih}
\phi \simeq \begin{cases} 
h &  \hspace{5mm} \textrm{for} \hspace{5mm} \phi<\phi_{\rm C}\,,  \\
\frac{M_P}{2\sqrt{\vert a\vert}} \log\left(1+\frac{h^2}{F_\infty^2}\right)& \hspace{5mm} \textrm{for} \hspace{5mm} \phi>\phi_{\rm C}\,,
   \end{cases}
\end{equation}
separated by a critical value
\begin{equation}\label{phiC}
 \phi_{\rm C}\equiv \frac{2M_P(1-6\vert a\vert)}{\sqrt{\vert a\vert }}\,.
\end{equation} 
The comparison between these approximate expressions and the exact field redefinition in Eq.~\eqref{phihexact} is shown in Fig.~\ref{fig:comparison}.

The large hierarchy between the transition scale $\phi_C$ and the electroweak scale allows us to identify in practice the vacuum expectation value $v_{EW}$ with $\phi=0$. In this limit, the Einstein-frame potential \eqref{potE} can be rewriten as 
 \begin{equation}\label{Vtree}
 V(\phi)\simeq \frac{\lambda}{4}F^4(\phi)\,,
 \end{equation}
 with 
  \begin{equation}\label{Ftree}
 F(\phi)\equiv 
 %\frac{h(\phi)}{\Omega(\phi)}\simeq
\begin{cases} 
\phi &  \hspace{2mm} \textrm{for} \hspace{2mm} \phi< \phi_{\rm C}\,, \\
 F_\infty \left(1-e^{-\frac{2\sqrt{\vert a\vert}\phi}{M_P}}\right)^{\frac12} 
 &  \hspace{2mm} \textrm{for} \hspace{2mm}\phi> \phi_{\rm C}\,.
   \end{cases}
 \end{equation} 
At $\phi<\phi_C$ we recover the usual Higgs potential (up to highly suppressed corrections, cf. Section \ref{sec:cutoff}).  At $\phi>\phi_C$ the Einstein-frame potential becomes exponentially stretched and approaches the asymptotic value $F_\infty$ at $\phi> M_P/(2\sqrt{\vert a\vert})$. 
The presence of $M_P$ in Eq.~\eqref{SHG} modifies also the decoupling properties of the Higgs field as compared to those in the induced gravity scenario. In particular, the masses of the intermediate gauge bosons and fermions in the Einstein-frame, ,\footnote{Here we use a compact notation for  the gauge boson couplings, namely  $g=g_2$ and $g_2 \cos\theta_w$ for the $B=W^\pm$ and $Z$ bosons respectively, with  $g_1$ and $g_2$ the gauge couplings of the $U(1)_Y$ and $SU(2)_L$ SM groups and $\theta_w=\tan^{-1}(g_1/g_2)$ the weak mixing angle. The coupling $y$ denotes a generic Yukawa coupling.}
\begin{equation}\label{masses}
m^2_B(\phi)  \equiv \frac{g^2}{4}  F^2(\phi)\,, \hspace{20mm} m_F(\phi)\equiv \frac{y}{\sqrt{2}} F(\phi)\,,
\end{equation}
coincide with the SM masses in the small field regime ($\phi<\phi_C$) and evolve towards constant values proportional to $F_\infty$ in the large-field regime ($\phi> M_P/(2\sqrt{\vert a\vert})$). 
The transition to the Einstein-frame effectively replaces $h$ by $F(\phi)$ in all (non-derivative) SM interactions. This behavior allows us to describe the Einstein-frame matter sector in terms of a chiral SM with vacuum expectation value $F(\phi)$ \cite{Bezrukov:2009db,Dutta:2007st}. 

\subsection{Tree-level inflationary predictions}\label{sec:predictionstree}
The flattening of the Einstein-frame potential \eqref{Vtree} due to the $\Theta=0$ pole allows for inflation with the usual slow-roll conditions even if the potential $V(\Theta)$ is not sufficiently flat. Let us compute the inflationary observables in the corresponding region $\phi>\phi_C$, where
 \begin{equation}\label{Vtree2}
V(\phi)\simeq \frac{\lambda F_\infty^4}{4} \left(1-e^{-\frac{2\sqrt{\vert a\vert}\phi}{M_P}}\right)^{2} \,.
 \end{equation}
The statistical information of the primordial curvature fluctuations generated by a single-field model like the one under consideration is mainly encoded in the two-point correlation functions of scalar and tensor perturbations, or equivalently in their Fourier transform, the power spectra. Following the standard approach \cite{Mukhanov:1990me}, we parametrize these spectra in an almost scale-invariant form,
\begin{equation} \label{defspec}
P_s=A_s\left(\frac{k}{k_*}\right)^{n_s-1}\,,\hspace{20mm} 
P_t=A_t\left(\frac{k}{k_*}\right)^{n_t}\,,
\end{equation}
and compute the inflationary observables
\begin{equation}\label{obsdef}
 A_s=\frac{1}{24\pi^2 M_P^4}\frac{V}{\epsilon}\,,  \hspace{15mm} n_s=1+ 2\eta-6\epsilon\,, \hspace{15mm}  r\equiv \frac{A_t}{A_s}=-8n_t=16\epsilon \,,
\end{equation}
with 
\begin{equation}\label{epsdef}
\epsilon\equiv\frac{M_P^2}{2}\left(\frac{V'}{V}\right)^2
%=\frac{8\vert a\vert }{\left(e^{2\sqrt{\vert a\vert}\phi_*/M_P}-1\right)^2}
\,, \hspace{25mm}
\eta \equiv M_P^2 \frac{V''}{V}\,,
%=\frac{8\vert a\vert \left(2-e^{2\sqrt{\vert a\vert}\phi_*/M_P}\right)}{\left(e^{2\sqrt{\vert a\vert}\phi_*/M_P}-1\right)^2} 
%\,,  \hspace{15mm}
%\delta^2 \equiv 
%\frac{M_P^4 V_{,{\phi}}V_{,\phi\phi\phi}}{ V^2}
%=\frac{64\vert a\vert^2 \left(e^{2\sqrt{\vert a\vert}\phi_*/M_P}-4\right)}{\left(e^{2\sqrt{\vert a\vert}\phi_*/M_P}-1\right)^3}\,,
\end{equation}
the first and second slow-roll parameters and the primes denoting derivatives with respect to $\phi$. The quantities in \eqref{obsdef} should be understood as evaluated at a field value $\phi_*\equiv \phi(N_*)$, with 
\begin{equation}\label{Nefolds}
N_*=\frac{1}{M_P}\int^{\phi_*}_{\phi_{\rm E}} \frac{d \phi}{\sqrt{2\epsilon}} =\left. \frac{1}{8\vert a\vert}\left(e^{2\sqrt{\vert a\vert}\phi/M_P}-\frac{2\sqrt{\vert a\vert}\phi}{M_P}\right)\right\vert_{\phi_{\rm E}}^{\phi_*}
\end{equation}
the e-fold number at which the reference scale $k_*$ in Eq.~\eqref{defspec} exits the horizon, i.e. $k_*=a_* H_*$. Here,
\begin{equation}
\phi_{\rm E}=\frac{M_P}{2\sqrt{\vert a\vert}}\ln\left(1+2\sqrt{2\vert a\vert}\right)\,,
\end{equation} 
stands for the field value at the end of inflation, which is defined, as usual, by the condition  $\epsilon(\phi_{\rm E})\equiv 1$. Equation \eqref{Nefolds} admits an exact inversion, 
\be \label{eq:inversion}
e^{2\sqrt{\vert a\vert}\phi_*/M_P}=-{\cal W}_{-1}\left[-e^{-8 \vert a\vert \bar N_*}\right]\,,
\ee
with ${\cal W}_{-1}$ the lower branch of the Lambert function and 
\be\label{Ninversion}
\bar N_* \equiv N_*+\frac{1}{8\vert a\vert}\left(e^{2\sqrt{\vert a\vert}\phi_E/M_P}-\frac{2\sqrt{\vert a\vert}\phi_E}{M_P}\right)\,,
%\bar N_* \equiv N_*+\frac{3}{4}\left[1+2\sqrt{2\vert a\vert}-\log \left(1+2\sqrt{2\vert a\vert}\right)\right]\,.
\ee
a rescaled number of e-folds. Inserting Eq.~\eqref{eq:inversion} into \eqref{obsdef} we get the following analytical expressions for the primordial scalar amplitude, 
\begin{equation}\label{eq:As}
A_s
 =\frac{\lambda(1-6\vert a\vert)^2}{12\,\pi^2\,\vert a\vert}\frac{(1+{\cal W}_{-1})^4}{ (8\vert a\vert {\cal W}_{-1})^2}  \,, 
\end{equation} 
its spectral tilt,
\begin{equation}\label{eq:ns}
n_s =1-16\vert a\vert  \frac{1-{\cal W}_{-1}}{\left(1+{\cal W}_{-1}\right)^2}\,, 
 %\hspace{20mm}
%\alpha_s = -128\,\vert a\vert ^2\frac{{\cal W}^2_{-1}-3{\cal W}_{-1}}{(1+{\cal W}_{-1})^4} \,.
\end{equation} 
and the tensor-to-scalar ratio
\begin{equation}\label{eq:r}
 r=\frac{128\, \vert a \vert }{\left(1+{\cal W}_{-1}\right)^2} \,.
\end{equation} 
%In the small $\vert a\vert$ limit, the inflationary predictions coincide with those of the $m^2\phi^2$ chaotic inflation scenario,
%\begin{equation}\label{appsmallk}
%n_s\simeq 1-\frac{4}{1+2N_*}\,, \hspace{25mm} r\simeq \frac{16}{1+2N_*}\,.
%\end{equation}
At large $\vert a\vert N_*$, these predictions display an interesting attractor behavior, very similar to that appearing in $\alpha$-attractor scenarios \cite{Ferrara:2013rsa,Kallosh:2013yoa,Galante:2014ifa} (see also Ref.~\cite{Artymowski:2016pjz}). Indeed, by taking into account the lower bound on the Lambert
function \cite{DBLP:journals/corr/Chatzigeorgiou16},
\begin{equation}
{\cal W}_{-1}[-e^{-8\vert a\vert \bar N_*}]>-8\vert a\vert \bar N_*-\sqrt{2(8\vert a\vert \bar N_*-1)} 
\,,
\end{equation}
we can obtain the approximate expressions\footnote{Note that the expressions contain $\bar N_*$ rather than $N_*$.}
\begin{equation}\label{applargek}
%A_s\simeq \frac{\lambda(1-6\vert a\vert)^2 \bar N_*^2 }{12 \pi ^2 \vert a\vert} \hspace{20mm}
n_s\simeq 1-\frac{2}{\bar N_*}\,, \hspace{25mm } r\simeq \frac{2}{ \vert a\vert \bar N_*^2}\,. 
%\hspace{20mm} \alpha_s=-\frac{2}{\bar N_*^2} \,.
\end{equation}
at $8\vert a\vert \bar N_*\gg 1$.
The free parameter $\vert a \vert$ (or equivalently the non-minimal coupling $\xi$) can be fixed by combining Eq.~\eqref{eq:As} with the normalization of the primordial spectrum at large scales \cite{Akrami:2018odb}, \begin{equation}\label{COBE}
\log (10^{10} A_s) \simeq 3.094 \pm 0.034\,.
\end{equation}
Doing this, we get a relation 
\begin{equation}\label{COBErel}
\xi \simeq 800 \bar N_* \sqrt{\lambda}\,, 
\end{equation}
among the non-minimal coupling $\xi$, the number of e-folds $\bar N_*$ and the Higgs self-coupling $\lambda$. 

The precise value of the number of e-folds in Eqs.~\eqref{applargek} and \eqref{COBErel} depends on the whole post-inflationary expansion and, in particular, on the duration of the heating stage. As the strength of the interactions among the Higgs field and the SM particles is experimentally known, the entropy production following the end of inflation can be computed in detail~\cite{GarciaBellido:2008ab,Bezrukov:2008ut,Repond:2016sol}.\footnote{This allows, for instance, to distinguish Higgs inflation from $R^2$ Starobinsky inflation \cite{Starobinsky:1980te,Bezrukov:2011gp}.} The depletion of the Higgs-condensate is dominated by  the non-perturbative production of massive intermediate gauge bosons, which, contrary to the SM fermions, can experience bosonic amplification. Once created, the $W^\pm$ and $Z$ bosons can decay into lighter SM fermions with a decay probability proportional to the instantaneous expectation value of the Higgs field $\phi(t)$. The onset of the radiation-domination era is determined either by i) the time at which the Higgs  amplitude approaches the critical value $\phi_C$ where the effective potential becomes quartic or by ii) the moment at which the energy density into relativistic fermions approaches that of the Higgs  condensate; whatever happens first. The estimates in Refs.~\cite{GarciaBellido:2008ab,Bezrukov:2008ut,Repond:2016sol} provide a range 
\begin{equation}
10^{13}\, {\rm GeV}\, \lesssim \, T_H\, \lesssim \, 2\times 10^{14}\, {\rm GeV} \,,
\end{equation}
with the lower and upper bounds associated respectively with the cases i) and ii) above. For the upper limit of this narrow window, we have $\bar N_*\simeq  N_*\simeq 59$ and we can rewrite Eq.~\eqref{COBErel} as a relation between $\xi$ and $\lambda$,
\begin{equation}
\xi \simeq 47200 \sqrt{\lambda}\,.
\end{equation}
Note that a variation of the Higgs self-coupling in this equation can be compensated by a change of the \textit{a priori} unknown non-minimal coupling to gravity. For the tree-level value $\lambda\sim {\cal O}(1)$, the non-minimal coupling must be significantly larger than one, but still much smaller than the value $\xi\sim  M^2_P/v^2_{\rm EW}\sim 10^{32}$ leading to sizable modifications of the effective Newton constant at low energies. In this regime, the parameter $\vert a\vert$ is very close to its maximum value $1/6$. This effective limit simplifies considerably the expression for the critical scale $\phi_C$ separating the low- and high-energy regimes,
\be 
\phi_C\simeq \sqrt{\frac{2}{3}}\frac{M_P}{\xi}\,,
\ee 
and collapses the inflationary predictions to the attractor values \cite{Bezrukov:2007ep}  
\begin{equation}\label{applargeknum}
%A_s\simeq \frac{\lambda(1-6\vert a\vert)^2 \bar N_*^2 }{12 \pi ^2 \vert a\vert} \hspace{20mm}
n_s\simeq 1-\frac{2}{\bar N_*}\simeq 0.966\,, \hspace{25mm } r\simeq \frac{12}{\bar N_*^2}\simeq 0.0034\,,
%\hspace{20mm} \alpha_s=-\frac{2}{\bar N_*^2} \,.
\end{equation}
in very good agreement with the latest results of the Planck collaboration \cite{Akrami:2018odb}. Note that, although computed in the Einstein frame, these predictions could have been alternatively obtained in the non-minimally coupled frame \eqref{lagr}, provided a suitable redefinition of the slow-roll parameters in order to account for the Weyl factor relating the two frames \cite{Makino:1991sg,Fakir:1992cg, Komatsu:1999mt,Tsujikawa:2004my, Koh:2005qp,Chiba:2008ia,Weenink:2010rr, Flanagan:2004bz, Chiba:2013mha,Postma:2014vaa, Jarv:2014hma,Ren:2014sya,Jarv:2015kga, Kuusk:2015dda,Jarv:2016sow, Burns:2016ric,Karam:2017zno,Karamitsos:2017elm, Karamitsos:2018lur}.

\section{Effective field theory interpretation} \label{sec:EFT}

The presence of gravity  makes Higgs inflation perturbatively non-renormalizable  \cite{Barbon:2009ya,Burgess:2009ea,Burgess:2010zq,Bezrukov:2010jz} and forbids its interpretation as an ultraviolet complete theory. The model should be therefore understood as an effective description valid up to a given cut-off scale $\Lambda$ \cite{Bezrukov:2010jz,George:2015nza}. This cutoff could either indicate the onset of a strongly coupled regime to be studied within the model by non-perturbative techniques (such as resummations, lattice simulations or functional renormalization studies) \cite{Aydemir:2012nz,Calmet:2013hia,Saltas:2015vsc,Escriva:2016cwl} or the appearance of new degrees of freedom beyond the initially-assumed SM content \cite{Giudice:2010ka,Barbon:2015fla}. 

\subsection{The cutoff scale}\label{sec:cutoff}

\textit{A priori}, the cutoff scale of Higgs inflation could coincide with the Planck scale, where gravitational effects should definitely taken into account.  Although quite natural, the identification of these two energy scales may not be theoretically consistent, since other  interactions could lead to violations of tree-level unitarity at a lower energy scale. An estimate\footnote{This procedure does not take into account possible cancellations among scattering diagrams, as those taking place, for instance, in models involving a \textit{singlet} scalar field not minimally coupled to gravity \cite{Hertzberg:2010dc}.} of the cutoff scale can be obtained by expanding the fields around their background values, such that all kind of higher dimensional operators appear in the resulting action \cite{Bezrukov:2010jz,Ferrara:2010in}. The computation is technically simpler in the original frame \eqref{SHG}. In order to illustrate the procedure let us consider the graviscalar sector in Eq.~\eqref{lagr}. Expanding the fields around their background values $\bar g_{\mu\nu}$ and $\bar h$,
\begin{equation}
  \label{gexp}
g_{\mu\nu} = \bar g_{\mu\nu}+\gamma_{\mu\nu}\,,
 \hspace{20mm}  
 h= \bar h+\delta h\,,
  \end{equation}
we obtain the following quadratic Lagrangian density for the perturbations $\gamma_{\mu\nu}$ and $\delta h$
\begin{eqnarray}
  \label{Lagr2}
    {\cal L}^{(2)} &=&
    \frac{M_P^2+\xi\bar h^2}{8}\big(\gamma^{\mu\nu}\Box \gamma_{\mu\nu}
    +2\partial_\nu \gamma^{\mu\nu}\partial^\rho \gamma_{\mu\rho}-2\partial_\nu \gamma^{\mu\nu}\partial_\mu \gamma
    -\gamma\Box \gamma\big) \nonumber \\
    &-&\frac{1}{2}(\partial_\mu\delta h)^2
    +\xi\bar h\big(
   \partial_\lambda\partial_\rho \gamma^{\lambda\rho}- \Box \gamma
    \big)\,\delta h
    \;,
\end{eqnarray}
with $\gamma=\bar g^{\mu\nu}\gamma_{\mu\nu}$ denoting the trace of the metric excitations. For non-vanishing $\xi$, the last term in this equation mixes the trace of the metric perturbation with the scalar perturbation $\delta h$ \cite{Barvinsky:2008ia,DeSimone:2008ei,Barvinsky:2009fy,Barvinsky:2009ii}.  To identify the different cutoff scales one must first diagonalize the kinetic terms. This can be done by performing a redefinition of the perturbations $(\gamma_{\mu\nu},\delta h)\to (\hat \gamma_{\mu\nu},\delta \hat h)$ with 
\begin{align}
  \label{hsubs}
  \gamma_{\mu\nu} &= \frac{1}{\sqrt{M^2_P+\xi\bar h^2}} \,\hat \gamma_{\mu\nu}
  -\frac{2\xi\bar h  \bar{g}_{\mu\nu}}{\sqrt{(M^2_P+\xi\bar h^2)(M^2_P+(1+
      6\xi)\xi\bar h^2)}} \,\,\delta\hat h\,, \\
  \label{chisubs}
  \delta h &=
  \sqrt{\frac{M^2_P+\xi\bar h^2}{M^2_P+(1+6\xi)\xi \bar h^2}}
  \,\delta\hat h
  \;.
\end{align}
Once Eq.~\eqref{Lagr2} has been reduced to a diagonal form, we can proceed to read the cutoff scales. The easiest one to identify is that associated with purely gravitational interactions,
\be\label{cutoff1}
\Lambda_{\rm P}(\bar h)\equiv \sqrt{M^2_P+\xi\bar h^2}\,,
\ee 
 which coincides with the effective Planck scale in Eq.~\eqref{lagr}.
For scalar-graviton interactions, the leading-order higher-dimensional operator is $(\delta\hat h)^2 \Box\hat \gamma/\Lambda_{\rm S}(\bar h)$, where
\be 
  \label{cutoff2}
  \Lambda_{\rm S}(\bar h) \equiv  \frac{M^2_P+(1+
    6\xi)\xi \bar h^2}{\xi\sqrt{M^2_P+\xi\bar h^2}} \;.
\ee 
Although we have focused on the graviscalar sector of the 
theory, the lack of renormalizability associated with the non-minimal coupling to gravity permeates all SM sectors involving the Higgs field. 
One could study, for instance, the scattering of intermediate $W^\pm$ and $Z$ bosons. Since we are working in the unitary gauge, it is sufficient to consider the longitudinal polarization. The modification of the Higgs kinetic term at large field values changes the delicate pattern of cancellations in the SM and leads to a tree-level unitarity violation at a scale
\begin{equation}\label{cutoff3}
\Lambda_{\rm G}(\bar h)\equiv \frac{\sqrt{M_P^2+\xi(1+6\xi)\bar h^2}}{\sqrt{6}\xi}\,.
\end{equation}
Note that the above scales depend on the background field $ \bar h$. For small field values ($\bar  h \lesssim M_P/\xi$), the cutoffs \eqref{cutoff1}, \eqref{cutoff2} and \eqref{cutoff3}
coincide with those obtained by naively expanding the theory around the electroweak scale, namely $\Lambda_{\rm P}\simeq M_P$, $\Lambda_{\rm S}\simeq  \sqrt{6}\Lambda_{\rm G}\simeq M_P/\xi $ \cite{Burgess:2009ea,Barbon:2009ya,Burgess:2010zq,Hertzberg:2010dc,Atkins:2010yg}. At large field values, ($\bar  h \gtrsim M_P/\xi$), 
the suppression scale depends on the particular process under consideration. For $M_P/\xi\ll\bar h\ll M_P/\sqrt{\xi}$ the graviscalar cutoff $\Lambda_{\rm S}$ grows quadratically 
till $\bar h\simeq M_P/\sqrt{\xi}$, where it becomes linear in $\bar h$ and traces the \textit{dynamical} Planck mass in that regime, $\Lambda_{\rm P}\simeq  \sqrt{\xi} \bar  h$. On the other hand, the gauge cutoff $\Lambda_{\rm G}$
 smoothly interpolates between $\Lambda_{\rm G}\sim  M_P/\xi$ at $\bar  h \lesssim M_P/\xi$ and $\Lambda_{\rm G}\sim g\bar h$ at  $\bar  h \gtrsim M_P/\xi$. Note that all cutoffs scales become linear in $\bar h$ at $\bar  h \gtrsim M_P/\xi$. This means that any operator $\Delta {\cal L}$ constructed out of them, the Higgs field and some Wilson coefficients $c_n$  approaches a scale-invariant form at large field values, namely
\begin{eqnarray}
\Delta {\cal L} &\equiv& \sum_n \hskip 2pt \frac{ c_n \hskip 2pt {\cal O}_n[\bar h]}{\left[\Lambda(\bar h)\right]^{n -4}} \simeq \sum_n \hskip 2pt \frac{ c_n \hskip 2pt {\cal O}_n[\bar h]}{(\sqrt{\xi} \bar h)^{n -4}}\sim \sum_n \hskip 2pt \frac{c_n  }{(\sqrt{\xi})^{n -4}} h^4\,, \hspace{20mm} n>4\,.
\end{eqnarray}

\subsection{Relation between high- and low-energy parameters}  

In what follows we will assume that the ultraviolet completion of the theory respects the original symmetries of the tree-level action, and in particular the approximate scale invariance of Eq.~\eqref{lagr} in the large-field regime and the associated shift-symmetry of its Einstein-frame formulation. This strong assumption forbids the generation of dangerous higher-dimensional operators that would completely spoil the predictivity of the model. In some sense, this requirement is not very different from the one implicitly assumed in other inflationary models involving trans-Planckian field displacements. 

The minimal set of higher-dimensional operators to be included on top of the tree-level action is the one generated by the theory itself via radiative corrections  \cite{Bezrukov:2010jz,Bezrukov:2014ipa}. The cancellation of the loop divergences stemming from the original action requires the inclusion of an infinite set of  counterterms with a very specific structure.  As in any other \textit{non-renormalizable} theory, the outcome of this subtraction procedure \textit{depends} on the renormalization scheme, with different choices corresponding to different assumptions about the ultraviolet completion of the theory.  Among the different subtractions setups, a dimensional regularization scheme involving a \textit{field-dependent} subtraction point \cite{Bezrukov:2009db}
\begin{equation}
\mu^2\propto M_P^2+\xi h^2\,,
\end{equation}
 fits pretty well with the approximate scale-symmetry of Eq.~\eqref{SHG} at large-field values.\footnote{The use of other schemes such as Pauli-Villars regularization or standard dimensional regularization with \textit{field-independent} subtraction point leads to dilatation-symmetry breaking and the consequent bending of the Higgs inflation plateau due to radiative corrections, see for instance Refs.~\cite{Barvinsky:2008ia, DeSimone:2008ei,Barvinsky:2009fy,Barvinsky:2009ii}.} Given this frame and scheme, the minimal set of higher-dimensional operators generated by the theory can be computed in any Weyl-related frame provided that all fields and dimensionfull parameters are appropriately rescaled. The computation becomes particularly simple in the Einstein-frame, where the Weyl-rescaled renormalization point $\mu^2 \Theta$ coincides with the standard field-independent prescription of renormalizable field theories, $\mu^2 \Theta\propto M_P^2$. 
A general counter-term in dimensional regularization contains a finite part $\delta {\cal L}$ and a divergent part in the form of a pole in $\epsilon=(4-d)/2$, with $d$ the dimension of spacetime. The coefficient of the pole is chosen to cancel the loop divergences stemming from the original action.  Once this divergent part is removed, we are left with the finite contribution $\delta {\cal L}$. The strength of this term encodes the remnants of a particular ultraviolet completion and cannot be determined within the effective field theory approach \cite{Bezrukov:2010jz,Bezrukov:2014ipa,Burgess:2014lza}. 
From a quantitative  point of view, the most relevant $\delta {\cal L}$ terms are related to the Higgs and top-quark interactions. In the Einstein-frame at one loop, they take the form ~\cite{Bezrukov:2014ipa}
\begin{equation}\label{counterL}
  \delta\cL^{F}_\text{1} =\left[\delta\lambda_a \left(F'^2+\frac{1}{3}F''F\right)^2 -\delta\lambda_b\right] F^4\,, \hspace{10mm}
  \delta\cL^{\psi}_\text{1} =\left[\delta y_{a} F'^2F  +\delta y_{b} F''(F^4)''\right]\bar\psi\psi\,,
\end{equation}
where the primes denote again derivatives with respect to $\phi$.  Note that these operators 
differ, as expected, from those appearing in the tree-level action. This means that, while the  contribution $\delta\lambda_b$ can be removed by a self-coupling redefinition, the finite parts $\delta\lambda_a, \delta y_{a}$  and $\delta y_{b}$ should be promoted to new couplings constants. Once the associated operators are added to the tree-level action, the re-evaluation of radiative corrections will generate additional contributions beyond the original one-loop result. These contributions come together with new finite parts that must be again promoted to novel couplings with their own renormalization group equations.
The iteration of this scheme leads to a renormalized action including an infinite set of higher-dimensional operators constructed out of the function $F$ and its derivatives. For small field values, the function $F$ becomes approximately linear ($F \approx \phi$, $F'=1$) and one recovers the SM non-minimally coupled to gravity up to highly suppressed interactions. In this limit, the coefficients of the infinite set of counterterms can be eliminated 
by a redefinition of the low energy couplings, as happens in a renormalizable theory. When evolving towards the inflationary region,  the function $F$ becomes approximately constant ($F_\infty =F_\infty$, $F'=0$) and some of the previously absorbed finite parts are dynamically subtracted. The unknown finite parts modify therefore the running of the SM couplings at the \textit{transition region} $\phi_{\rm C}<\phi<\sqrt{3/2} \,M_P$, such that the SM masses at the electroweak scale \textit{cannot}  be unambiguously related to their inflationary counterparts \textit{without a precise knowledge of the ultraviolet completion} \cite{Bezrukov:2014ipa,Burgess:2014lza,Hertzberg:2011rc}. 

If the finite contributions are of the same order as the loops generating them, the  tower of higher dimensional operators generated by radiative corrections can be truncated  \cite{Bezrukov:2014ipa}. In this case, the effect of the 1-loop threshold corrections can be imitated by an effective change\footnote{This replacement implicitly neglects the running of the finite parts $\delta \lambda_a$ and $\delta y_a$ in the transition region $\phi_{\rm C}<\phi<\sqrt{3/2} \,M_P$. }  \cite{Bezrukov:2014ipa}
\begin{equation}
  \label{lambdajump}
  \lambda(\mu) \to \lambda(\mu)
  + \delta\lambda_a \left[
    \left(F'^2+\frac{1}{3}F''F\right)^2-1
    \right]\,, \hspace{10mm}  
  y_t(\mu) \to y_t(\mu)
  + \delta y_a \left[F'^2-1\right]\,,
\end{equation}
with $\lambda(\mu)$ and $y_t(\mu)$ given by the SM renormalization group equations. We emphasize, however, that the truncation of the renormalization group equations is not essential for most of the results presented below, since,  within the self-consistent approach to Higgs inflation, the functional form of the effective action is almost insensitive to it ~\cite{Bezrukov:2010jz,Bezrukov:2014ipa,Bezrukov:2017dyv}. 
%Within the minimalistic approach to Higgs inflation, the threshold effects affect only the shape of the \textit{transition region} $\phi_{\rm C}<\phi<\sqrt{3/2} \,M_P$. 

\subsection{Potential scenarios and inflationary predictions}\label{sec:predictionsEFT}

To describe the impact of radiative corrections on the inflationary predictions, we will make use of the renormalization group enhanced potential. This is given by the one in Eq.~\eqref{Vtree2} but with the Higgs self-coupling $\lambda$ replaced by its corresponding running value $\lambda(\phi)$,
\begin{equation}\label{Veff}
V(\phi)=\frac{\lambda (\phi)}{4}F^4(\phi)\,. 
\end{equation}
Note that we are not promoting the non-minimal coupling $\xi$ within $F(\phi)$ to a running coupling $\xi(\phi)$---as done, for instance, in Ref.~\cite{Ezquiaga:2017fvi}---but rather assuming it to be constant during inflation. This is indeed a reasonable approximation since the  one-loop beta function determining the running of $\xi$  \cite{Bezrukov:2009db,Yoon:1996yr},
\begin{equation}
 \beta_\xi(\mu) = \mu \frac{\partial}{\partial \mu} \xi = - \frac{1}{16 \pi^2} \xi \left( \frac{3}{2} g'^2 + 3 g^2 - 6 y_t^2\right)\,,
\end{equation}
is rather small for realistic values of the couplings constant at the inflationary scale, $\beta_\xi \propto \mathcal{O}(10^{-2})$ \cite{Bezrukov:2017dyv,Masina:2018ejw} (see also Ref.~\cite{Salvio:2017oyf}).

Although, strictly speaking, the renormalization group enhanced potential is not gauge invariant, the gauge dependence is small during slow-roll inflation, especially in the presence of extrema \cite{Espinosa:2015qea,Espinosa:2016nld,Cook:2014dga}. In the vicinity of the minimum of $\lambda(\phi)$, we can use the approximation \cite{Bezrukov:2014bra}
\begin{equation}\label{lambdaeff}
\lambda(\phi) =\lambda_0+b \log^2 \left(\frac{m_t(\phi)}{q}\right)\,,
\end{equation}
with the parameters $\lambda_0$, $q$ and $b$ depending on the \textit{inflationary values} of the  \textit{Einstein-frame} Higgs and top quark masses, according to the fitting formulas \cite{Bezrukov:2014bra}
\begin{eqnarray}\label{lambdazeroandb}
  \lambda_0 &=&\ 0.003297 \left[(m_H^*-126.13) 
               - 2 (m_t ^*-171.5) \right]\,, \notag\\
  q &=& \ 0.3 M_P \exp \left[]0.5(m_H^*-126.13)-0.03(m_t^*-171.5) \right]\,,
\\
 \nonumber
  b &= &\ 0.00002292 - 1.12524\times10^{-6}\left[ (m_H^*-126.13) -
 1.75912 (m_t^*-171.5)\right]\,, 
\end{eqnarray}
with  $m_H^*$ and $m_t^*$ in GeV.  As seen in the last expression, the parameter $b$, standing for the derivative of the beta function for $\lambda$ at the scale of inflation, is rather insensitive to the Higgs and top quark mass values at that scale and can be well-approximated by $b\simeq2.3\times10^{-5}$.  The choice 
\begin{equation}\label{muscale}
\frac{m_t(\phi)}{q}= \alpha \cdot \frac{y_t}{\sqrt{2}}\frac{F(\phi)}{q}\equiv \frac{\sqrt{\xi}F(\phi)}{\kappa M_P}\,,
\end{equation}
with $\alpha=0.6$ optimizes the convergence of perturbation theory \cite{Bezrukov:2009db,Bezrukov:2008ej}, while respecting the asymptotic symmetry of the tree-level action \eqref{lagr} and its non-linear shift-symmetric Einstein-frame realization. In the second equality, we have introduced an effective parameter $\kappa$ to facilitate the numerical computation of the inflationary observables. 

  %%%%%%%%%%%%%%%%%%%%%%%%%%%%%
\begin{figure}
\centering
\includegraphics[width=0.43\textwidth]{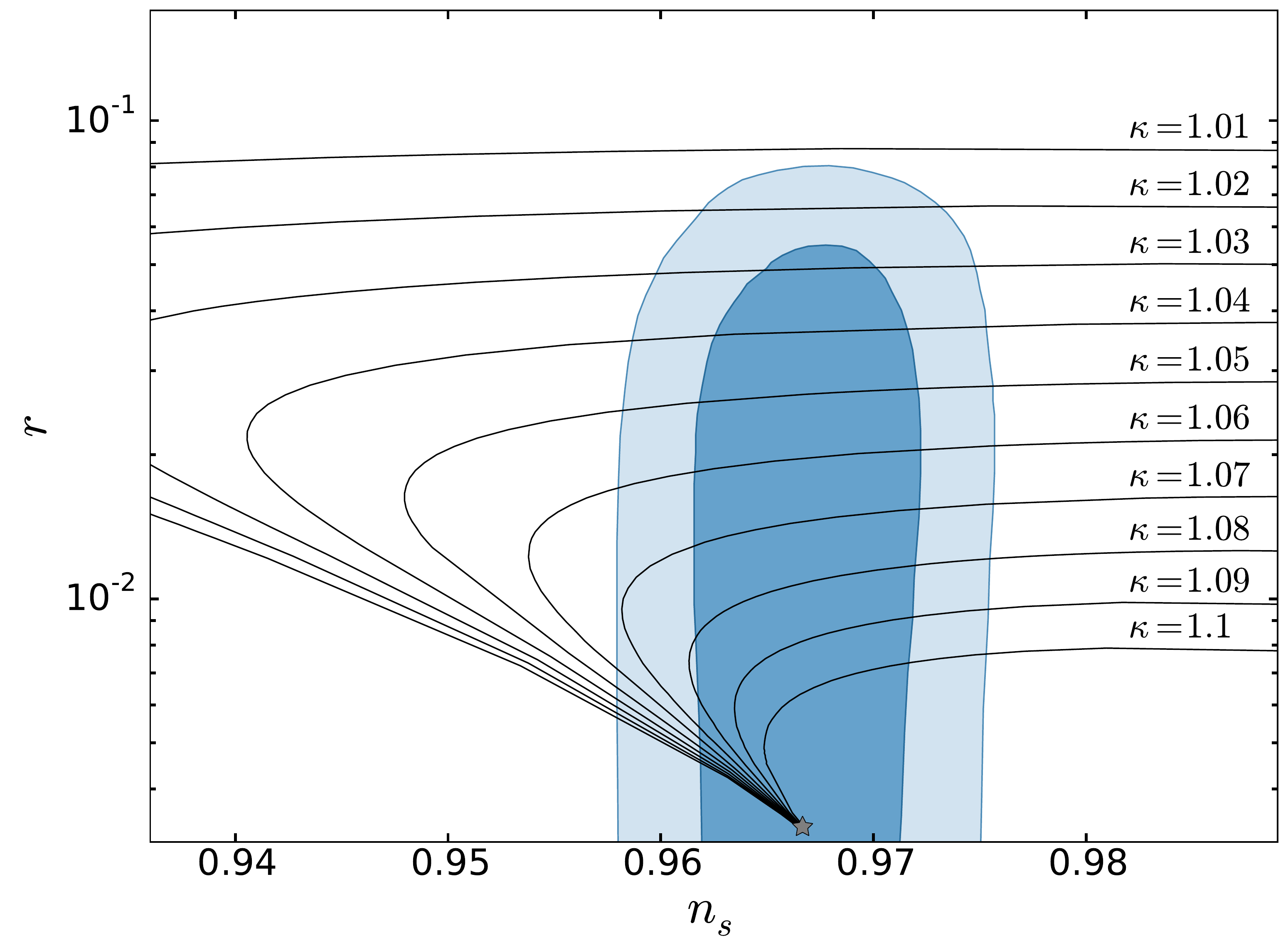}
\includegraphics[width=0.42\textwidth]{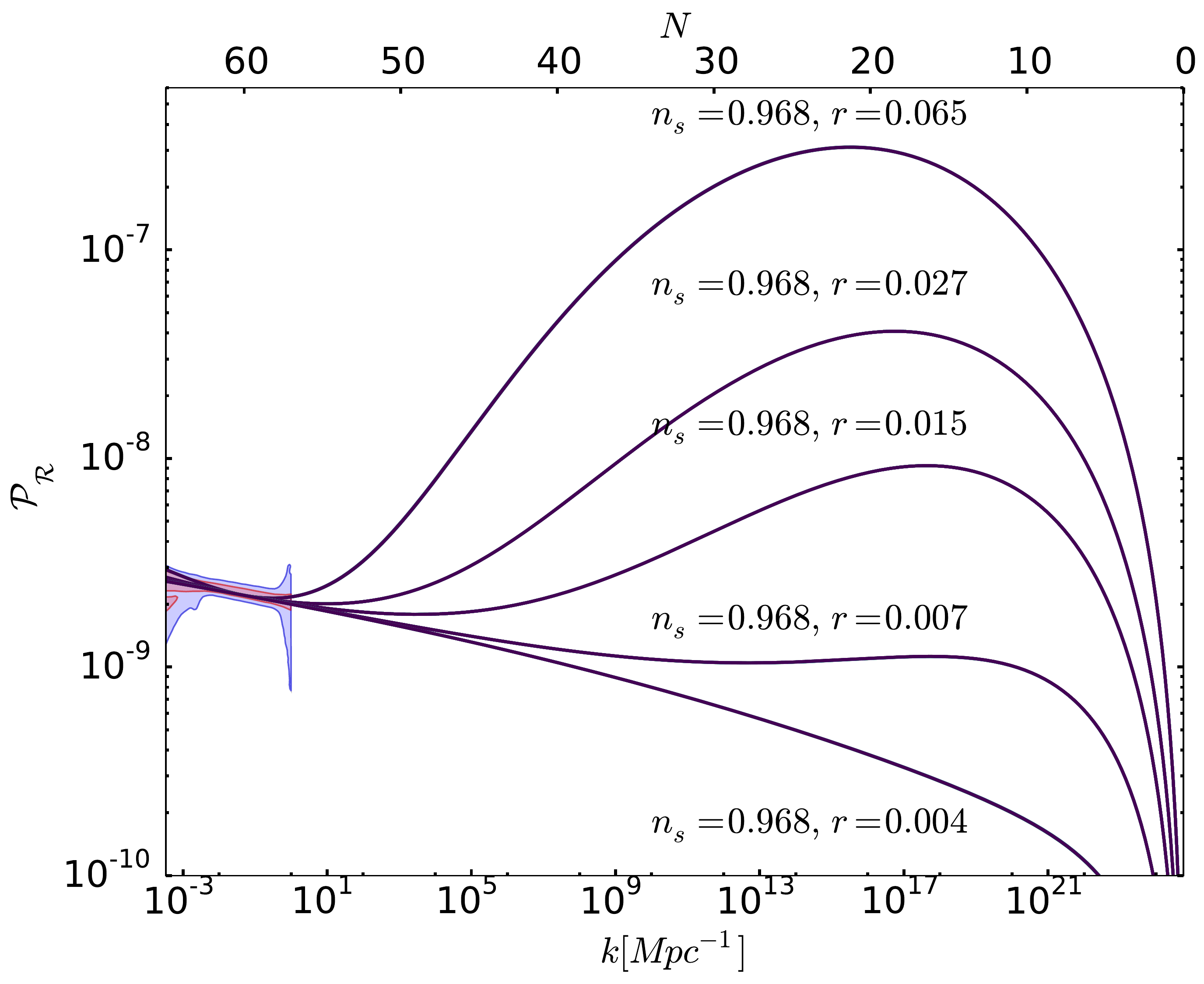}
\caption{(Left) The tensor-to-scalar ratio $r$ and the spectral tilt $n_S$ following from the effective potential \eqref{Veff} \cite{Bezrukov:2017dyv}. The non-minimal coupling $\xi$ varies between $10$ and $100$ along the lines of constant $\kappa$, with larger values corresponding to smaller tensor-to-scalar ratios.  The star in the lower part of the plot stands for the universal values in Eq.~\eqref{applargeknum}. The blue contours indicate the latest 68\% and 95\% C.L. Planck constraints on the $r$-$n_s$ plane \cite{Akrami:2018odb}.
(Right) The power spectrum ${\cal P}_{\cal R}$ as a function of the number of e-folds before the end of inflation and the associated comoving scale $k$ in inverse megaparsecs \cite{Bezrukov:2017dyv}. The monotonic curve at the bottom of the plot corresponds to the \textit{universal/non-critical Higgs inflation} scenario. The upper non-monotonic curves are associated with different realizations of the \textit{critical Higgs inflation scenario}.  The shaded regions  stand for the latest 68\% and 95\% C.L. constraints provided by the Planck collaboration \cite{Akrami:2018odb}.}\label{spectrumplot}
\end{figure}

A simple inspection of Eqs.~\eqref{Veff} and \eqref{lambdaeff} allows us to distinguish three regimes: 
\begin{enumerate}[i)]

\item {\it Non-critical regime/Universal}: 
If $\lambda_0 \gg b/(16\kappa)$, the effective potential \eqref{Veff} is almost independent of the radiative logarithmic correction and can be well approximated by its tree-level form \eqref{Vtree2}. Consequently, the inflationary observables retain their tree-level values \cite{Fumagalli:2016lls,Enckell:2016xse,Bezrukov:2014bra,Bezrukov:2017dyv}, cf.~Fig.~\ref{spectrumplot}. 

\item {\it Critical regime}:  If $\lambda_0 \gtrsim  b/(16\kappa)$, 
the first two derivatives of the potential are approximately zero, $V'\simeq V''\simeq 0$, leading to the appearance of a \textit{quasi}-inflection point at  
\begin{equation}\label{phiinf}
\phi_{\rm I}=\sqrt{\frac{3}{2}}\log\left(\frac{\sqrt{e}}{\sqrt{e}-1}\right)M_P\,.
\end{equation}
Qualitatively, the vast majority of inflationary e-folds in this scenario takes place in the vicinity of the inflection point $\phi_{\rm I}$, while the inflationary observables depend on the form of the potential as some value $\phi_*>\phi_I$.  

Given the small value of the Higgs self-coupling in this scenario, $\lambda_0\sim {\cal O}(10^{-6})$, the nonminimal coupling $\xi$ can be significantly smaller than in the universal regime, $\xi\sim{\cal O}(10)$, while still satisfying the normalization condition \eqref{COBE}  \cite{Allison:2013uaa,Bezrukov:2014bra,Hamada:2014iga,Hamada:2014wna}. This drastic decrease of the non-minimal coupling alleviates the tree-level unitary problems discussed in Section \ref{sec:cutoff} by raising the cutoff scale.

For small $\xi$ values, the tensor-to-scalar ratio can be rather large, $r\sim {\cal O}(10^{-1})$ \cite{Allison:2013uaa,Bezrukov:2014bra,Hamada:2014iga,Hamada:2014wna} (see also Ref.~\cite{Masina:2018ejw}). Note, however, that although CMB data seems to be consistent with the primordial power spectrum at large scales, the simple expansion in \eqref{defspec} cannot  accurately describe its global behavior since the running of the spectral tilt $\alpha_s\equiv d \ln n_s/d\ln k$ and its scale dependence $\beta_s\equiv d^2\ln n_s/d \ln k^2$ also become considerably large, cf.~Fig.~\ref{fig:critrunning}.  

The non-monotonic evolution of the slow-roll parameter $\epsilon$ in the vicinity of the inflection point leads to the enhancement of the spectrum of primordial density fluctuations at small and intermediate scales. It is important to notice at this point that the standard slow-roll condition may break down if the potential becomes extremely flat and the inertial contribution in the equation of motion for the inflation field is not negligible as compared with the Hubble friction \cite{Kannike:2017bxn,Germani:2017bcs,Garcia-Bellido:2017mdw}. In this regime, even the classical treatment is compromised since stochastic effects can no longer be ignored \cite{Starobinsky:1994bd,Vennin:2015hra,Pattison:2017mbe,Ezquiaga:2018gbw}.
%%%%%%%%%%%%%%%%%%
\begin{figure}
\begin{center}
\includegraphics[width=0.42\textwidth]{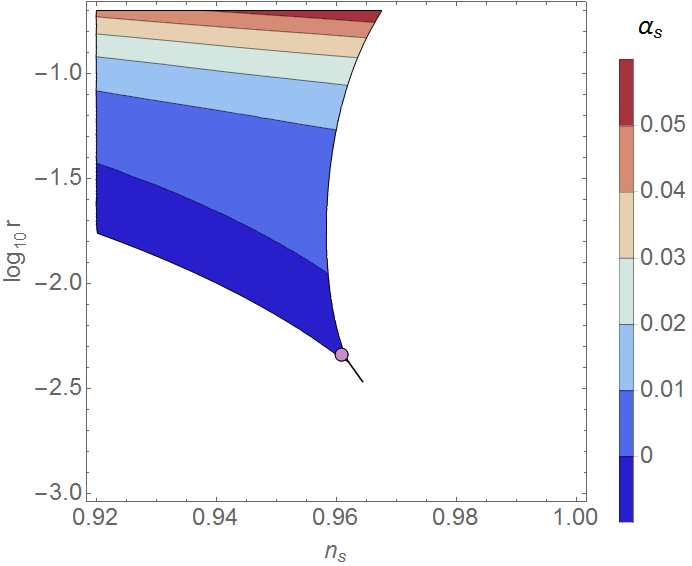}
\includegraphics[width=0.42\textwidth]{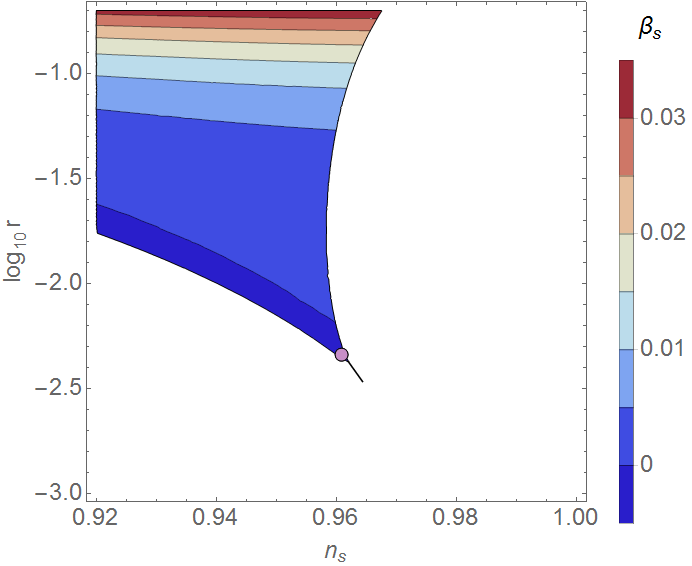}
\end{center}
\caption{(Left) Spectral tilt running  $\alpha_s\equiv d \ln n_s/d\ln k$ in critical Higgs inflation as a function of the tensor-to-scalar ratio $r$ and the spectral-tilt $n_s$ ~\cite{Rasanen:2017ivk}. (Right) Scale dependence of the spectral-tilt  $\beta_s\equiv d^2\ln n_s/d \ln k^2$ in the same case ~\cite{Rasanen:2017ivk}. The purple dots indicate the universal/non-critical Higgs inflation regime. The boundaries on the right-hand side of the figures correspond to the constraint on the number of e-folds following the heating estimates in Refs.~\cite{GarciaBellido:2008ab,Bezrukov:2008ut,Repond:2016sol}. For lower heating efficiencies, the boundaries move to the left, decreasing the spectral tilt but not significantly affecting the tensor-to-scalar ratio \cite{Rasanen:2017ivk}.}\label{fig:critrunning}
\end{figure}
%%%%%%%%%%%%%%%%%%

If we restrict ourselves to situation in which the slow-roll approximation is satisfied during the whole inflationary trajectory,\cite{Bezrukov:2017dyv}\footnote{The onset of the slow-roll regime prior to the arrival of the field  to the inflection point and its dependence on pre-inflationary conditions was studied in Ref.~\cite{Salvio:2017oyf}, where a robust inflationary attractor was shown to exist.} the height and width of the generated bump at fixed spectral tilt are correlated with the tensor-to-scalar ratio $r$, cf.~Fig.~\ref{spectrumplot}. Contrary to some claims in the literature \cite{Ezquiaga:2017fvi}, the maximum amplitude of the power-spectrum compatible  with the 95\% C.L Planck $n_s-r$ contours \cite{Bezrukov:2017dyv} is well below the critical threshold ${\cal P}^{\rm max}_{\cal R}\simeq  10^{-2}-10^{-3}$ needed for primordial black hole formation \cite{Carr:2016drx, Bird:2016dcv,Carr:2017jsz} (see, however, Refs.~\cite{Ezquiaga:2018gbw} and \cite{Rasanen:2018fom}). This conclusion is unchanged if one considers the effect of non-instantaneous threshold corrections \cite{Bezrukov:2017dyv}, which could potentially affect the results given the numerical proximity of the inflection point $\eqref{phiinf}$ to the upper boundary of the transition region, $\phi\simeq \sqrt{3/2}M_P$.

\item \textit{Hilltop regime}: If $\lambda\lesssim b/(16\kappa)$ the potential develops a new minimum at large field values \cite{Fumagalli:2016sof,Rasanen:2017ivk}. This minimum is separated from the electroweak minimum by a local maximum where hilltop inflation can take place \cite{Boubekeur:2005zm,Barenboim:2016mmw}. This scenario is highly sensitive to the initial conditions since the inflaton field must start on the electroweak vacuum side and close enough to the local maximum in order to support an extended inflationary epoch. On top of that, the fitting formulas in \eqref{lambdazeroandb} may not be accurate enough for this case, since they are based on an optimization procedure around the $\lambda(\phi)$ minimum. The tensor-to-scalar ratio in this scenario differs also from the universal/non-critical Higgs inflation regime, but contrary to the critical case, it is decreased to $2\times 10^{-5}<r < 1\times 10^{-3}$, rather than increased \cite{Rasanen:2017ivk,Fumagalli:2016lls}.
\end{enumerate}

\subsection{Vacuum metastability and high-temperature effects}\label{sec:vacuuminsta} 

\begin{figure}
\begin{center}
\includegraphics[width=0.43\textwidth]{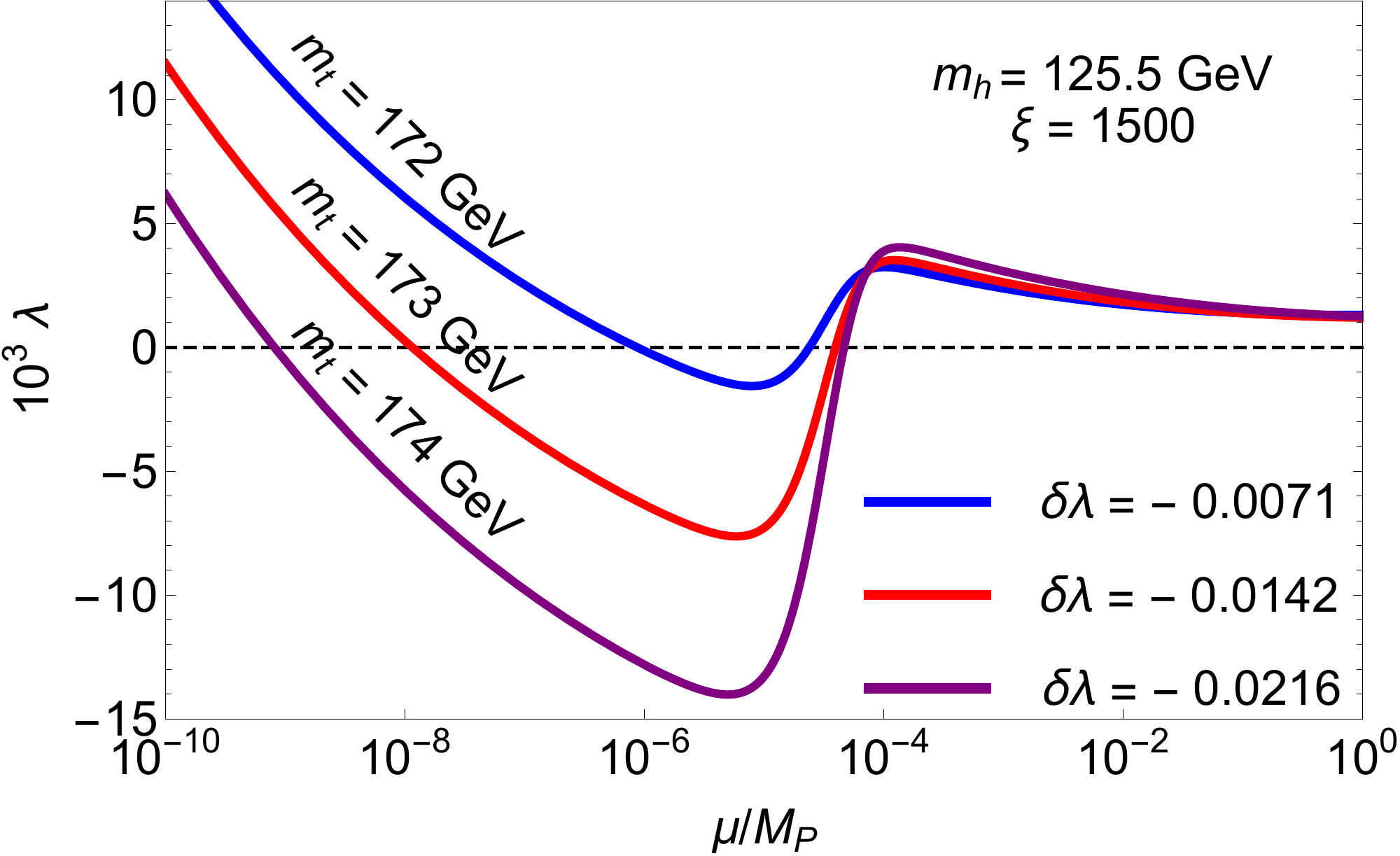}
\includegraphics[width=0.42\textwidth]{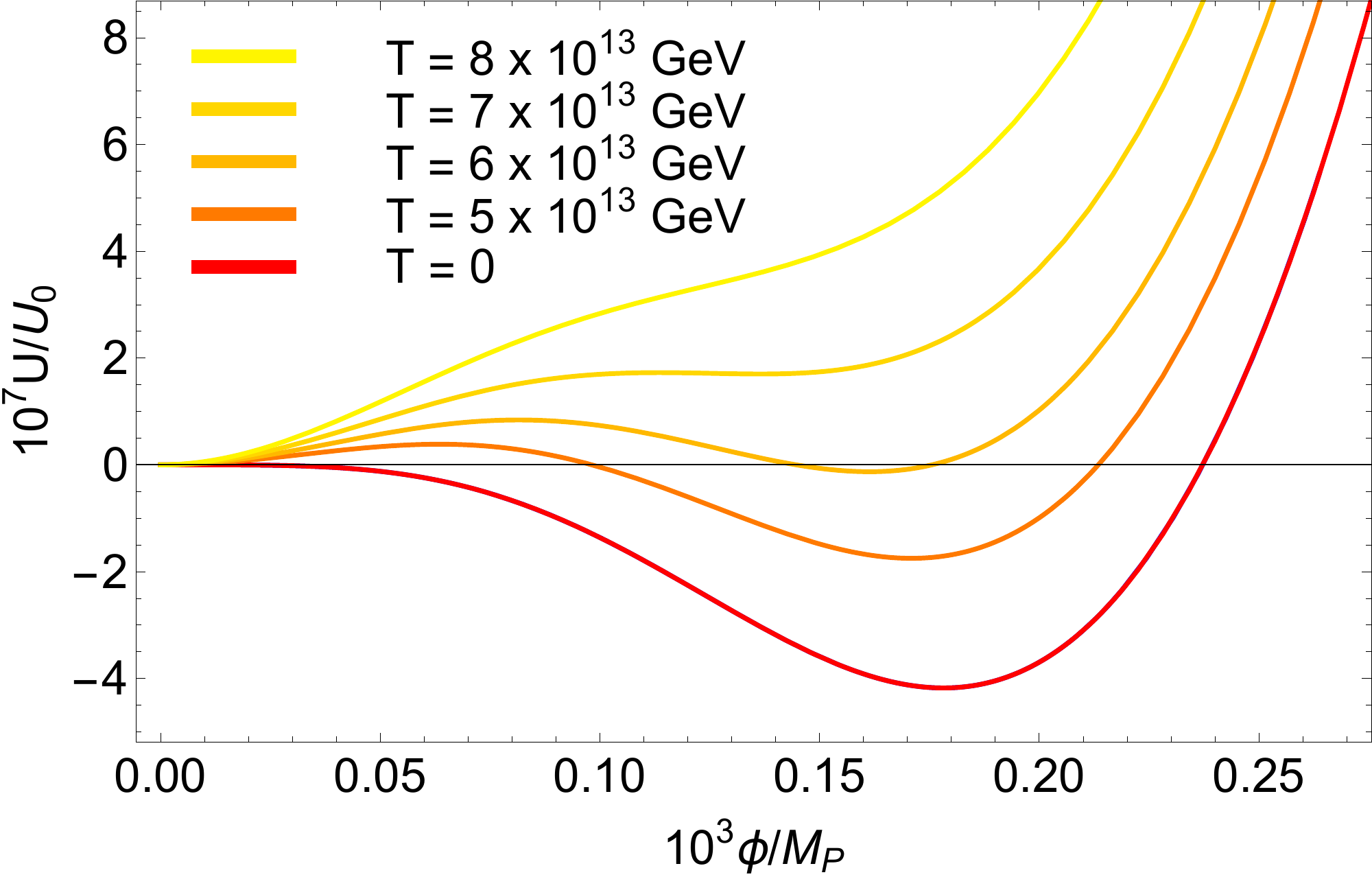}
\end{center}
\caption{(Left) Illustrative values of the 1-loop threshold corrections needed to restore the asymptotic behavior of the  universal Higgs inflation potential at large field values for electroweak SM pole masses leading to SM vacuum instability at $\mu_0\sim 10^{9}, 10^{10}$ and $10^{12}$ GeV. (Right) Comparison between the potential following from the set of parameters leading to the red line in the previous plot and the thermally-corrected effective potential accounting for the backreaction effects of the decay products created during the heating stage. The normalization factor $U_0=(10^{-3} M_P)^4$ account for the typical energy density at the end of inflation. }\label{fig:restoration}
\end{figure}

The qualitative classification of scenarios and predictions  presented in the previous section depends on the  \textit{inflationary values} of the Higgs and top quark masses and holds independently of the value of their electroweak counterparts. In particular, any pair of couplings following from the SM renormalization group equations can be connected to a well-behaved pair of couplings in the chiral phase by a proper choice of the unknown threshold corrections. This applies also if the SM vacuum is not completely stable. Some examples of the 1-loop threshold correction $\delta \lambda_a$ needed to restore the universal/non-critical Higgs inflation scenario beyond $\mu_0\sim 10^{9}, 10^{10}$ and $10^{12}$ GeV are shown in Fig.~\ref{fig:restoration}.  For a detailed scan of the parameter space see Refs.~\cite{Fumagalli:2016lls,Enckell:2016xse}.
 
 %%%%%%%%%%%%%%%%%%
%%%%%%%%%%%%%%%%%%

The non-trivial interplay between vacuum stability and threshold corrections generates an additional minimum at large field values. Provided the usual chaotic initial conditions, the Higgs field will evolve in the trans-Planckian field regime,  inflating the Universe while moving towards smaller field values. Since the new minimum is significantly wider and deeper than the electroweak one, it seems likely that the Higgs field will finish its post-inflationary evolution there. Note, however, that this conclusion is strongly dependent  on the  ratio of the Higgs energy density to the second minimum depth. If this ratio is large, the entropy production at the end of inflation may significantly modify the shape of the potential, triggering its stabilization and allowing the Higgs field to evolve towards the desired electroweak vacuum \cite{Bezrukov:2008ut}. 

The one-loop finite temperature corrections to be added on top of the Einstein-frame renormalization group enhanced potential take the form \cite{Linde:1978px}
\be\label{thermal}
\Delta V= T \sum_{i=B,F} \int\frac{d^3k}{(2\pi)^3 a^3}{\rm ln}\left[1\pm \exp\left(-\frac{k^2/a^2 + m_i^2}{T}\right)\right]\,,
\ee
with the plus and minus signs corresponding respectively to fermions and bosons and $m_{B,F}$ standing for the Einstein-frame masses in Eq.~\eqref{masses}. The most important contributions in Eq.~\eqref{thermal} are associated with the top quark and the electroweak bosons, with the corresponding coupling constants $y_t$ and  $g$ evaluated at $\mu_{y_t}=1.8\, T$ and $\mu_g=7\, T$,  in order to minimize the radiative corrections \cite{Kajantie:1995dw}.  

A detailed analysis of the universal/non critical Higgs inflation scenario reveals that the temperature of the decay products generated during the heating stage exceeds generically the temperature at which the unwanted secondary vacuum at large field values disappears \cite{GarciaBellido:2008ab,Bezrukov:2008ut}, see Fig~\ref{fig:restoration}.\footnote{A detailed scan of the parameter space \textit{assuming instantaneous conversion of the inflaton energy density into a thermal bath} was performed in Ref.~\cite{Enckell:2016xse}.} The stabilization becomes favored for increasing $\mu_0$ values\footnote{The larger $\mu_0$ is, the shallower and narrower the ``wrong" minimum becomes, cf. Fig.~\ref{fig:restoration}.} and holds even if this scale is as low as $10^{10}$ GeV \cite{Bezrukov:2008ut}. The thermally-corrected potential enables the Higgs field to relax to the SM vacuum.  After the heating stage, the temperature decreases as the Universe expands and the secondary minimum reappears, first as a local minimum and eventually as the global one. When that happens, the Higgs field is already trapped in the electroweak vacuum. Although  the barrier separating the two minima  prevents a direct decay, the Higgs field could still tunnel to the global minimum. The probability for this to happen is, however, very small and the lifetime of SM vacuum significantly exceeds the life of the Universe \cite{Anderson:1990aa,Arnold:1991cv,Espinosa:1995se,Espinosa:2007qp}. 
Universal/non-critical Higgs inflation with a graceful exit can therefore take place \textit{for electroweak SM pole masses leading to vacuum metastability at energies below the inflationary scale.} \cite{Bezrukov:2008ut}. 

The situation changes completely if one considers the critical Higgs inflation scenario. In this case, the energy of the Higgs condensate is comparable to the depth of the secondary minimum and symmetry restoration does not take place. Unless the initial conditions are extremely fine-tuned, the Higgs field will relax to the minimum of the potential at Planckian 
values, leading with it to the inevitably collapse of the Universe \cite{Felder:2002jk}. The success of critical Higgs inflation requires therefore the absolute stability of the electroweak vacuum  \cite{Bezrukov:2008ut}.

\section{Variations and extensions}\label{sec:variations}
 
Many variations and extensions of Higgs inflation have been considered in the literature, see for instance Refs. \cite{Lerner:2010mq,Giudice:2010ka,BenDayan:2010yz,Kamada:2010qe,Lerner:2011ge,Arai:2011nq,Steinwachs:2013tr,Kamada:2012se,Einhorn:2012ih,Greenwood:2012aj,Kanemura:2012ha,Choudhury:2013zna,Oda:2014rpa,Xianyu:2014eba,Oda:2014mza,Ellis:2014dxa,He:2014ora,Kamada:2015sca,Okada:2015zfa,Cai:2015gla,Lazarides:2015xia,vandeBruck:2015gjd,Takahashi:2016uyq,Calmet:2016fsr,Ellis:2016spb,Okada:2016ssd,Ge:2016xcq,Marian:2017bvz,Ema:2017ckf,Ema:2017rqn,He:2018gyf,Chen:2018ucf}. In what follows we will restrict ourselves to those proposals that are more closely related to the minimalistic spirit of the original scenario. In particular, we will address a Palatini formulation of Higgs inflation and the embedding of the model to a fully scale invariant framework.

\subsection{Palatini Higgs inflation} 

In the usual formulation of Higgs inflation presented in Section \ref{HI_Jordan}, the action  is minimized with respect to the metric. This procedure implicitly assumes the existence of a Levi-Civita connection depending on the metric tensor and the inclusion of a York-Hawking-Gibbons term ensuring the cancellation of a total derivative term with no-vanishing variation at the boundary \cite{York:1972sj,Gibbons:1976ue}. One could alternatively consider a Palatini formulation of gravity in which the metric tensor and the connection are treated as independent variables and no additional boundary term is required to obtain the equations of motion \cite{Ferraris1982}. Roughly speaking, this formulation corresponds to assuming an ultraviolet completion involving different gravitational degrees of freedom. 

Although the metric and Palatini formulations of General Relativity give rise to the  same equations of motion \cite{Ferraris1982}, this is not true for scalar-tensor theories as Higgs inflation. To see this explicitly let us consider the Higgs inflation action in Eq.~\eqref{lagr} with $R= g^{\mu\nu}R_{\mu\nu}(\Gamma,\partial \Gamma )$ and $\Gamma$ a
\textit{non}-Levi-Civita connection.\footnote{For a recent generalization built from the Higgs, the metric and the connection and involving only up to two derivatives see Ref.~\cite{Rasanen:2018ihz}.} Performing a Weyl rescaling of the metric $g_{\mu\nu} \to \Theta\, g_{\mu\nu}$ with $\Theta$ given by Eq.~\eqref{Thetadef} we obtain an Einstein-frame action 
\be \label{einsteinframe1}
S=\int d^4x \sqrt{-g}\left[\frac{M_P^2}{2}g^{\mu\nu}R_{\mu\nu}(\Gamma)-\frac{1}{2}M_P^2 K(\Theta) (\partial \Theta)^2 - V(\Theta)\right]  \,,
\ee
containing a potential \eqref{potE} and a non-canonical kinetic term with 
\be \label{K_HI_Pal}
K(\Theta)\equiv \frac{1}{4\vert a \vert \Theta^2}\left(\frac{1}{1-\Theta}\right)\,,
%\frac{M_{P}^2+\xi h^2}{\Omega^2(h)\left(M_{P}^2+\xi h^2\right)}\,, 
\ee
and
\be \label{ainfPala}
\vert a\vert \equiv \xi \,.
\ee 
Note that the kinetic function \eqref{K_HI_Pal} differs from that obtained in the metric formulation, see ~Eq.~\eqref{K_HI}. In particular, it does not contain the part associated with the non-homogeneous transformation of the Ricci scalar, since $R=R(\Gamma)$ is now invariant under Weyl rescalings. For the purposes of inflation, this translates into a modification of the residue of the inflationary pole at $\Theta=0$ with respect to the metric case. While the metric value of $\vert a\vert$ in Eq.~\eqref{K_HI} is bounded from above (cf.~Eq.~\eqref{ainf}), it can take positive arbitrary values in the Palatini formulation (cf.~Eq.~\eqref{ainfPala}). Performing a field redefinition
\begin{equation} \label{chi1}
	\frac{1}{M^2_P}\left(\frac{d\phi}{d\Theta}\right)^2 = K(\Theta)\hspace{10mm}\longrightarrow  \hspace{10mm}
    h(\phi)=F_\infty \sinh\left(\frac{\sqrt{a}\phi}{M_{P}}\right)\,,
\end{equation}
to canonically normalize the $h$-field kinetic term, we can rewrite the graviscalar action~\eqref{einsteinframe1} at $\phi\gg v_{\rm EW}$ as
\be \label{EframeS1}
S=\int d^4x \sqrt{-g}\left[\frac{M_P^2}{2}R-\frac{1}{2}(\partial \phi)^2 - V(\phi)\right]  \,, 
\ee
with 
\begin{equation} \label{chipotential1}
V(\phi) = \frac{\lambda}{4} F^4(\phi)\,, \hspace{20mm} F(\phi) \equiv
	F_\infty\tanh\left(\frac{\sqrt{a}\phi}{M_{P}}\right)\,.
\end{equation}
The comparison of the latest expression with Eq.~\eqref{Ftree} reveals 
some important differences between the metric and Palatini formulations. In both cases, the effective Einstein-frame potential smoothly interpolates between a low-energy quartic potential and an asymptotically flat potential at large field values. Note, however, that the transition in the Palatini case is rather direct and does not involve the quadratic piece appearing in the metric formulation. On top of that, the flatness of the asymptotic plateau is different in the two cases, due to the effective change in $\vert a\vert$. The Palatini dependence $\vert a\vert=\xi$ has a strong impact on the inflationary observables. In the large-field regime they read 
\begin{equation}\label{applargekPal}
n_s\simeq 1-\frac{2}{\bar N_*}\,, \hspace{25mm } r\simeq \frac{2}{\xi \bar N_*^2}\,, 
\end{equation}
with 
\be 
\bar N_*\equiv N_*+\frac{1}{16 \vert a \vert }\cosh\left(\frac{2\sqrt{a}\phi_E}{M_P} \right)\,, 
\ee 
a rescaled number of e-folds and 
\be \label{endofinf}
	\phi_{\rm E} = 
		\frac{M_P}{2\sqrt{a}}  \arcsinh(\sqrt{32 a}) \,,
\ee
the inflaton value at the end of inflation ($\epsilon(\phi_{\rm end})\equiv 1$), with $\phi_{\rm E}=\sqrt{3/2}\arcsinh(4/\sqrt{3})\,M_P $ corresponding to the $\xi\to \infty$ limit and $\phi_{\rm E}=2\sqrt{2}\,M_P$ to the end of inflation in a minimally coupled $\lambda \phi^4$ theory. A relation between the non-minimal coupling $\xi$, the self-coupling $\lambda$ and the number of e-folds $\bar N_*$ can be obtained by taking into account the amplitude of  the observed power spectrum in Eq.~\eqref{COBE},
\be \label{xicondition1}
\xi \simeq 	3.8\times10^6 \bar N_*^2\lambda\,.
\ee 
A simple inspection of Eq.~\eqref{applargekPal} reveals that the predicted tensor-to-scalar ratio in Palatini Higgs inflation is within the reach of current or future experiments \cite{Matsumura:2016sri} only if $\xi\lesssim10$, which, assuming $\bar N\simeq 59$, requires a very small coupling $\lambda \lesssim 10^{-9}$. For a discussion of unitarity violations in the Palatini formulation see Ref.~\cite{Bauer:2010jg}.

\subsection{Higgs-Dilaton model} \label{sec:HD}

The existence of robust predictions in (non-critical) Higgs inflation is intimately related to the emerging dilatation symmetry of its tree-level action at large field values. The uplifting of Higgs inflation to a completely scale-invariant setting was considered in Refs.~\cite{Shaposhnikov:2008xb,GarciaBellido:2011de,Blas:2011ac,Bezrukov:2012hx,GarciaBellido:2012zu,Rubio:2014wta,Trashorras:2016azl,Karananas:2016kyt,Ferreira:2016vsc,Ferreira:2016wem,Ferreira:2016kxi,Ferreira:2018itt,Casas:2017wjh,Ferreira:2018qss,Casas:2018fum}. 
In the unitary gauge $H=(0,h/\sqrt{2})^T$, the graviscalar sector of the Higgs-Dilaton model considered in these papers takes the form
\beq
\label{eqn:hdm_lagrangian}
S=\int d^4 x \sqrt{-g} \left[\frac{\xi_h h^2 +\xi_\chi \chi^2}{2}
R 
-\frac12 (\partial h)^2- \frac{1}{2} (\partial \chi)^2-V(h,\chi)\right]\,,
\eeq
with 
\beq \label{eq:pot}
 \hspace{5mm}U(h,\chi)=\frac{\lambda}{4} \left(h^2-\alpha\chi^2 \right)^2 + \beta \chi^4
\eeq
a scale-invariant version of the SM symmetry-breaking potential and $\alpha,\beta$ positive dimensionless parameters. The existence of a well-defined gravitational interactions at all field values requires the non-minimal gravitational couplings to be positive-definite, i.e. $\xi_h,\xi_\chi>0$. In the absence of gravity, the ground state of Eq.~\eqref{eqn:hdm_lagrangian} is determined by the scale-invariant potential \eqref{eq:pot}. For $\alpha\neq 0$ and $\beta=0$, the vacuum manifold extends along the flat directions $h_0=\pm \alpha \chi_0$. Any solution with $\chi_0\neq 0$ breaks scale symmetry spontaneously and induces non-zero values for the effective Planck mass and the 
electroweak scale.\footnote{
Among the possible values  of $\beta$ in the presence of gravity, the case $\beta=0$ seems also preferred \cite{GarciaBellido:2011de,Allen:1987tz,Jalmuzna:2011qw}, 
see also Refs.~\cite{Antoniadis:1985pj,Tsamis:1992sx,Tsamis:1994ca,Antoniadis:2006wq,Polyakov:2009nq,Wetterich:2017ixo}.} 
The relation between these highly hierarchical scales is set by fine-tuning $\alpha\sim v^2/M_P^2\sim 10^{-32}$. For this small value, the flat valleys in the potential $U(h,\chi)$ are essentially aligned and we can safely approximate $\alpha\simeq 0$ for all inflationary purposes.

To compare the inflationary predictions of this model with those of the standard Higgs-inflation scenario, let us perform a Weyl rescaling $  g_{\mu\nu} \rightarrow M_P^{2}/(\xi_h h^2 +\xi_\chi \chi^2)g_{\mu\nu}$ followed by a field redefinition \cite{Casas:2017wjh}
\be
\gamma^{-2} \Theta \equiv \frac{(1+6\xi_h)h^2+(1+6\xi_\chi)\chi^2}{\xi_h h^2+\xi_\chi\chi^2}\,,\label{eq:Thetadef} \hspace{10mm}
\exp\left[\frac{2\gamma\Phi}{M_P}\right] \equiv \frac{a}{\bar a} \frac{(1+6\xi_h)h^2+(1+6\xi_\chi)\chi^2}{M_P^2}\,,
\ee
with 
\beq\label{kappadef}
\gamma \equiv \sqrt{\frac{\xi_\chi}{1+6\xi_\chi}} \,, \hspace{15mm}
a \equiv-\frac{\xi_h}{1+6\xi_h}\,, \hspace{15mm} 
\bar a \equiv a \left(1-\frac{\xi_\chi}{\xi_h}\right)\,. 
\eeq
After some algebra, we obtain a rather simple Einstein-frame action \cite{Karananas:2016kyt,Casas:2017wjh}
\begin{equation}\label{action_HD2}
S=\int d^4 x\sqrt{-g}\left[  \frac{M_P^2}{2}R 
-\frac{1}{2}M_P^2 K(\Theta)(\partial\Theta)^{2} 
- \frac{1}{2}\Theta(\partial \Phi)^2 
 -U(\Theta)\right]\,,
\end{equation}
containing a potential  
\beq\label{Utheta}
U(\Theta)= U_0(1-\Theta)^2\,, \hspace{20mm} U_0\equiv\frac{\lambda \, M_P^4}{4}\left(\frac{1+6 \bar a}{\bar a}\right)^2\,,
\eeq
and a non-canonical, albeit diagonal, kinetic sector. The kinetic function for the $\Theta$ field,
\begin{equation}\label{poles} 
K(\Theta)=\frac{1}{4\,\vert \bar a \vert \Theta^2}\left(\frac{c}{\vert \bar a \vert\Theta-c}+\frac{1-6 \vert \bar a\vert \Theta}{1-\Theta}\right)\,,
\end{equation}
contains two ``inflationary" poles at $\Theta=0$ and $\Theta=c/\vert\bar a\vert$  and a ``Minkowski'' pole at $\Theta=1$, where the usual SM action is approximately recovered.  As in the single field case, the ``Minkowski'' pole does not play a significant role during inflation and can be neglected for all practical purposes. Interestingly, the field-derivative space becomes in this limit a maximally symmetric hyperbolic manifold with Gaussian curvature $a<0$ \cite{Karananas:2016kyt}.  

Inflation takes place in the vicinity of the inflationary poles. 
During this regime, the kinetic term of the $\Phi$-field is effectively suppressed and the dilaton rapidly approaches a constant value $\Phi=\Phi_0$. This effective freezing is an immediate consequence of scale invariance. As in the single field case, the shift symmetry $\Phi\to\Phi+C$ in Eq.~\eqref{action_HD2} allows us to interpret 
$\Phi$ as the dilaton or Goldstone of dilatations. As first shown  in Ref.~\cite{GarciaBellido:2011de} (see also Refs.~\cite{Ferreira:2016wem,Ferreira:2018itt}), the equation of motion for this field coincides with the scale-current conservation equation, effectively restricting the evolution to constant $\Phi$ ellipsoidal trajectories in the $\lbrace h,\chi\rbrace$ plane.   Given this emergent \textit{single-field} dynamics, no non-gaussianities nor isocurvature perturbations are significantly generated during inflation \cite{GarciaBellido:2011de,Ferreira:2018qss}. If the $\Theta$ variable is dominated by the Higgs component ($\xi_h\gg \xi_\chi$), the spectral tilt and  the tensor-to-scalar ratio take the compact form
\begin{equation}
n_s \simeq 1-8\,c \coth\left(4 c N_*\right)\,,  \hspace{20mm} r\simeq   \frac{32\, c^2}{\vert a \vert} \csch^{2} \left(4 c N_*\right)\,,
\label{nsr1} 
\end{equation}
with $\vert a \vert \simeq 1/6$ in order to satisfy the normalization condition \eqref{COBE}. 
Note that these expressions rapidly converge to the Higgs inflation values \eqref{applargek} for $4 c N_*\ll 1$. For increasing $c$ and fixed $N_*$, the spectral tilt decreases linearly and the tensor-to-scalar ratio approaches zero.

\section{Concluding remarks} 

Before the start of the LHC, it was widely believed that we would find a plethora of new particles and interactions that would reduce the Standard Model to a mere description of Nature at energies below the TeV scale. 
From a bottom-up perspective, new physics was typically advocated to cure the divergences associated with the potential growth of the Higgs self-coupling at high energies.  The finding of a relatively light Higgs boson in the Large Hadron Collider concluded the quest of the Standard Model spectrum while demystifying the concept of naturalness and the role of fundamental scalar fields in particle physics and cosmology. The Standard Model is now a confirmed theory that could stay valid till the Planck scale and provide a solid theoretical basis for describing the early Universe. 

The Higgs field itself could lead to inflation if a minimalistic coupling to the Ricci scalar is added to the Standard Model action. The value of this coupling can be fixed by the normalization of the spectrum of primordial density perturbations, leaving a theory with no free parameters at tree level. On top of that,  the experimental knowledge of the Standard Model couplings reduces the usual uncertainties associated with the heating stage and allows us to obtain precise predictions in excellent agreement with observations.  
Note, however, the mere existence of gravity makes the theory non-renormalizable and forces its interpretation as an effective field theory. Even in a self-consistent approach to Higgs inflation, the finite parts of the counterterms needed to make the theory finite obscure the connection between low- and high-energy observables. If these unknown coefficients are small, Higgs inflation provides an appealing relation between the Standard Model parameters and the properties of the Universe at large scales. If they are large, this connection is lost but Higgs inflation can surprisingly take place even when the Standard Model vacuum is not completely stable. 

\section{Acknowledgments}
The author acknowledges support from the Deutsche Forschungsgemeinschaft through the Open Access Publishing funding programme of the Baden-W\"urttemberg Ministry of Science, Research and Arts and the Ruprecht-Karls-Universit\"at Heidelberg as well as through the project TRR33 ``The Dark Universe''. He thanks Guillem Domenech, Georgios Karananas and Martin Pauly for useful comments and suggestions on the manuscript. 
 \vspace{1cm}
 
 \footnotesize
\begin{multicols}{2}
\small{
\bibliographystyle{unsrt}
\bibliography{biblio.bib}
}
\end{multicols}

\end{document}